\documentclass[fleqn,usenatbib]{mnras}

\usepackage{newtxtext,newtxmath}

\usepackage[T1]{fontenc}

\DeclareRobustCommand{\VAN}[3]{#2}
\let\VANthebibliography\thebibliography
\def\thebibliography{\DeclareRobustCommand{\VAN}[3]{##3}\VANthebibliography}

\usepackage{graphicx}	
\usepackage{amsmath}	

\usepackage{float}

\newcommand{\eq}[1]{Eq.~(\ref{#1})}

\title[The case of HD 106906]{The case of HD 106906 debris disc: A binary's revenge}
\author[M. Farhat, A. Sefilian, \& J. Touma]{
Mohammad A. Farhat,$^{2}$
Antranik A. Sefilian,$^{3,4,5}$\thanks{Alexander von Humboldt Fellow.}  
Jihad R. Touma$^{1, 4}$\thanks{Corresponding Author: jt00@aub.edu.lb}
\\
\\
$^{1}$Department of Physics, American University of Beirut
PO Box 11-0236, Riad El-Solh, Beirut 11097 2020, Lebanon
\\
$^{2}$IMCCE, CNRS, Observatoire de Paris, PSL University, Sorbonne Université, 77 Avenue Denfert-Rochereau, 75014, Paris, France
\\
$^{3}$Astrophysikalisches Institut und Universit{\"a}tssternwarte, Friedrich-Schiller-Universit\"at Jena, Schillerg\"a{\ss}chen~2--3, 07745 Jena, Germany
\\
$^{4}$Center for Advanced Mathematical Sciences, American University of Beirut, PO Box 11-0236, Riad El-Solh, Beirut 11097 2020, Lebanon
\\
$^{5}$Departamento de F{\'i}sica, Universidad T{\'e}cnica Federico Santa Mar{\'i}a, Av. España 1680, Valpara{\'i}so, Chile
}
\date{Accepted XXX. Received YYY; in original form ZZZ}

\pubyear{2022}

\begin{document}
\label{firstpage}
\pagerange{\pageref{firstpage}--\pageref{lastpage}}
\maketitle

\begin{abstract}
Debris disc architecture presents [exo-]planetary scientists with precious clues for processes
of planet formation and evolution, including constraints on planetary mass perturbers. This is particularly true of the disc in HD\,106906, which in early HST, then follow up polarimetric observations, presented asymmetries and needle-like features that have been attributed to perturbations by a massive, and unusually distant external planetary companion. Here, we revisit the long-term secular dynamical evolution of the HD\,106906 disc allowing for the combined gravitational action of the planetary companion and the inner stellar binary which holds the system together. We argue that the binary is strong enough to impose a dynamical break at the disc’s location, resulting in distinctive observational signatures which we render via simulated surface density maps and vertical structure profiles. Within uncertainties on the planet’s orbit, we show that the disc can go from being fully dominated by the inner binary to significantly so, and is hardly ever outside its reach. The extent of binary dominance impacts the disc’s mean eccentricity, a metric which we map as a function of the planet’s semi-major axis and orbital eccentricity, with and without radiation pressure. We can thus constrain the planet’s orbit to ease the tension between evident axisymmetry in the millimeter, and apparent asymmetry in scattered light. We discuss phase space structure, then inclination distribution, arguing for the relevance of our results to a variety of hierarchical systems, as we set the stage for generalizations that allow for disc self-gravity and collisional evolution.
\end{abstract}

\begin{keywords}
planet–disc interactions
 -- planets and satellites: dynamical evolution and stability -- circumstellar matter -- stars: individual: HD 106906
\end{keywords}



\section{Introduction}
HD 106906 features a 13 Myr-old tight spectroscopic stellar binary \citep{pecaut2012revised,de2019near}, a directly imaged planet \citep{bailey2013hd,daemgen2017high}, and a debris disc in between \citep{kalas2015direct,Lagrange-2016}. In recent years, improved constraints on the planet's mass and orbit were accompanied with attempts at understanding disc architecture (asymmetries and apparent eccentricity in HST then polarimetric observations; e.g., \citealp{crotts2021deep}), then implications for the origin of that planet in the first place \citep{Nguyen_2020}.

To be sure, planet mass and distance present challenges for run-of-the-mill planet formation scenarios, with theorists considering the by now familiar pathways around similar such conundrums: in-situ formation in the course of binary genesis, capture in the host stellar cluster, formation close-in then ejection via instability in a multi-planet system, or through direct mean-motion resonance between migrating planet and binary \citep[e.g.,][]{bate2002formation,Beust-2017,jennings2021primordial,Moore-et-al-2022}. An exploration of the latter scenario by \cite{Beust-2017} was alone in allowing for the stellar binary’s gravitational field, mainly on the migrating/ejected planet, but also on the debris disc in ways that presage our present work [a matter we shall come back to below]. 

Otherwise, all studies of debris dynamics in the HD 106906 disc to date have focused on the planet as the sole gravitational perturber \citep[e.g., ][]{jilkova2015debris,nesvold2016circumstellar,nesvold2017hd}, while ignoring the role of the binary in shaping the structure of that ring of debris, if not controlling it altogether. In other words, the inner binary present in HD 106906 was assumed sufficiently tightly bound for its quadrupole to be rather benign when compared to the planet's perturbations. This left the disc in the throngs of an eccentric Kozai-Lidov trap \citep{naoz2016eccentric,nesvold2017hd}, through which it was natural to obtain and argue for significant excitation of eccentricities and inclinations over the rather young age of the system (13 Myrs). This further appeared consistent with available observations of the disc's asymmetries \citep{kalas2015direct,crotts2021deep,van2021survey}, which more often than not were sensitive to dust and its multifarious perturbations, and were not particularly adept at detecting the underlying distribution of planetesimals, themselves the subject of modeling. 
\begin{figure*}
\centering
\includegraphics[width=.8\linewidth]{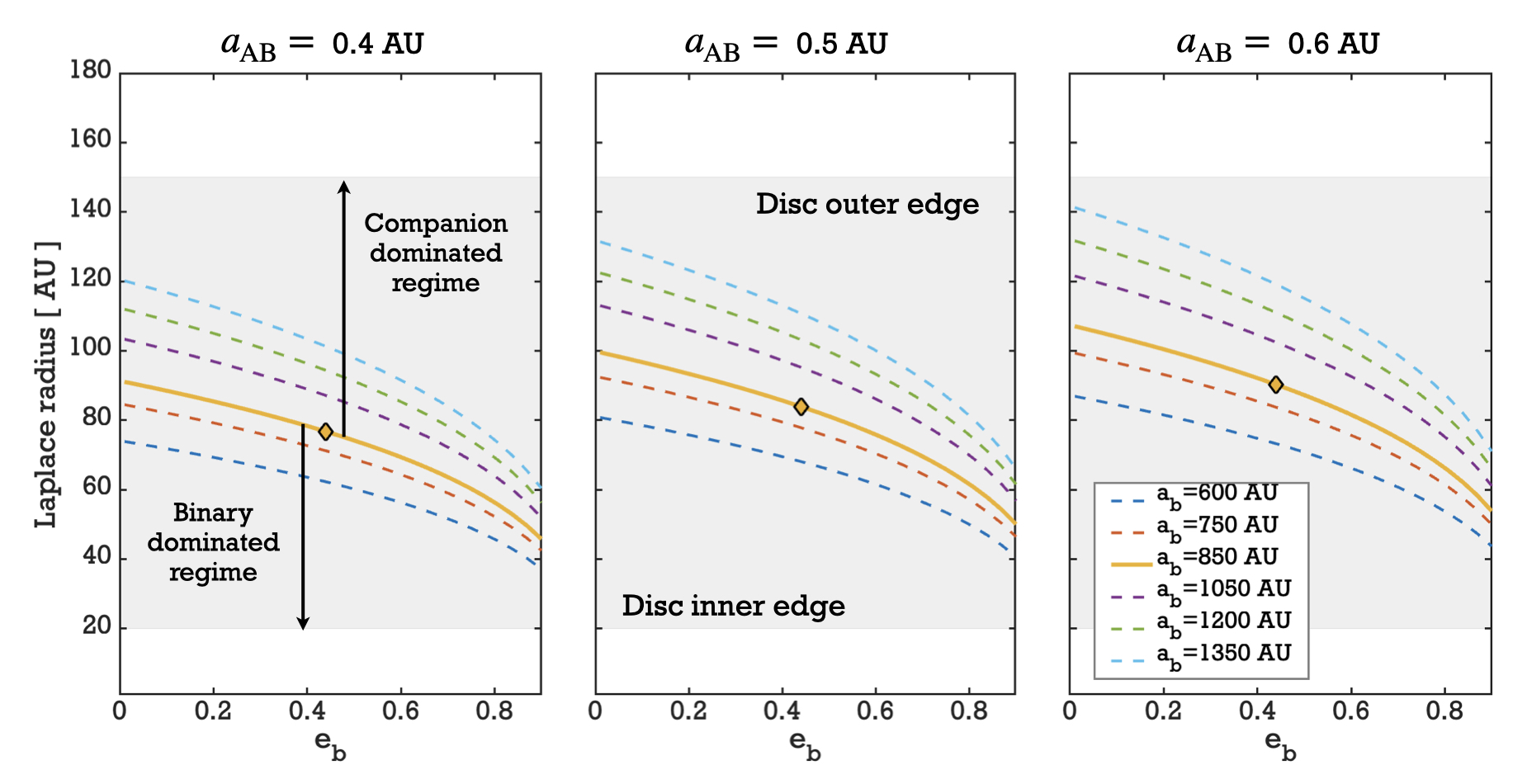}
\caption{Laplace Radius Revealed. Plotted versus planet eccentricity $e_{\rm b}$  is the location of the Laplace radius (Eq. \ref{laplace_radius}) in the HD 106906 system, computed for various values of the planetary semi-major axis as shown in the legend. Calculations are done for three different values of the stellar binary's semi-major axis as indicated on top of each panel. The variations are consistent with current observational constraints \citep[][see also Table \ref{table_system_parameters}]{Beust-2017,Nguyen_2020}. For all of these variations, the resulting curve is shown to fall within the range of semi-major axes associated with the debris disc (the gray highlighted region), and to separate it into two substantial chunks, one dominated by the inner binary and the other by the planetary companion. Markers correspond to the nominal values of the planetary parameters ($a_{\rm b}=850 {\rm AU},e_{\rm b}=0.44)$ reported in \citet{Nguyen_2020}.}
\label{fig:radius}
\end{figure*}

We are fresh out of a study of an intimately related system, the Solar system, where we investigated the fate of Trans-Neptunian debris in the combined field of the outer gaseous giants inside and a putative super-earth on a wide inclined and eccentric orbit outside \citep{farhat2021laplace}. In that setting, the inner and outer perturbers were recognized as comparable and shown to confine planetesimals into eccentric and inclined frozen orbits. It was thus natural for us to start by allowing for the inner binary in the HD 106906 system and learn in the process that the binary makes up in mass what it loses in proximity, rendering its quadrupolar action comparable to the wide planetary companion, HD 106906 b, over a substantial range of debris semi-major axes. 

Subjecting the debris disc to the combined gravitational influence of the inner binary and the outer planetary companion gives rise to the so-called Laplace surface. The latter, originally posited by \cite{laplace1805vol} in his classic study of Jupiter’s satellites, interpolates a family of planes parameterized by the satellite's orbital distance. On these planes, now referred to as the Laplace planes, competing torques from the Jovian bulge and the solar tide annihilate on stationary circular orbits with ``proper'' inclinations. Spatially, these planes coincide with the planet’s equator on the inside and tend to the planet’s orbit when moving outwards, smoothly forming the Laplace surface.

With a view to greater realism, \cite{Tremaine} relaxed the assumption of circular equilibria, and studied the stability of the classic Laplace surface to perturbations in the satellite's eccentricity and to variations in planetary obliquity. Their results ushered a stream of solar system and exo-planetary generalizations exploring:  the early evolution of the Moon around a fast spinning oblique Earth \citep{cuk2016tidal, tian2020vertical}, hemispherical asymmetries in Uranian satellites \citep{charnoz_canup_crida_dones_2018}, warps in circumplanetary discs \citep{zanazzi2016extended}, and non-gravitational perturbations such as radiation pressure on dust grains \citep{rosengren2014laplace, tamayo2013dynamical}. More recently, \cite{farhat2021laplace} generalized the framework in \cite{Tremaine} by allowing for symmetry breaking octupolar contributions from an eccentric external perturber. Such contributions were shown to destroy the classical Laplace surface of circular orbits, replacing it with an eccentric analogue together with bifurcating families of spatially frozen orbits at large inclination and eccentricity. 

In this work, we highlight the significance and draw the implications of the eccentric Laplace surface of \cite{farhat2021laplace} for debris disc dynamics in HD 106906. We start with estimates arguing for the applicability of this dynamics to this remarkable system.  We then proceed to deploy the relevant machinery on the debris disc, identifying equilibrium structures, then following dynamical evolution starting from an initially cold disc. We subject the evolving disc to both kinematic and photometric measures. We comment on disc properties of HD 106906's age, then consider long-term simulations and associated observational signatures with similar such systems in mind. Dynamical evolution is further situated in the appropriate phase space, arguing that a single geometric structure captures the full dynamics, as it prepares the stage for future collisional studies of debris discs around a Laplace surface. The work raises serious questions for HD 106906 debris modeling to date and should be eminently relevant for emerging ALMA observations. While deeply rooted in HD 106906 debris disc dynamics, our study is of broader relevance to discs around binaries with distant companions, and may once again feed into Planet 9 fantasies closer to home \citep[e.g.,][]{batygin2019planet}.  

\section{You may wish the binary away, but the disc knows otherwise: Critical spatio-temporal scales}
We are interested in studying the long-term dynamics of large , i.e., km-sized, planetesimals constituting the HD 106906 debris disc. Before delving  into detailed calculations, and mainly with fellow observers in mind, we start by spatially locating the HD 106906 debris disc with respect to the competing gravitational influences of the inner stellar binary (here captured to quadrupolar order) and the outer super-Jupiter. What we are inquiring about is of course germane to the Laplace surface associated with this system, and we shall introduce and map it with sufficient detail in what follows. For now, we simply point out that a critical role in this balancing act is played by a distinguished radius, the so called Laplace radius, $r_{\rm L}$ for short. This radius signals the transition from inner-binary to outer-companion dominated dynamics, a point of inflection in a family of equilibrium orbits interpolating between the inner binary's orbital plane and the plane of the outer companion's orbit. It appears as a dimensionless parameter controlling those equilibria, and in our particular setting takes the form
\begin{equation}\label{laplace_radius}
    r_{\rm L}^5 = a_{\rm b}^3 \,a_{\rm AB}^2 \frac{m_{\rm A} m_{\rm B}}{{m_{\rm AB}^2}}\frac{m_{\rm AB}}{m_{\rm b}} {[1- e_{\rm b}^2]}^\frac{3}{2},
\end{equation}
where $m_{\rm b}$, $a_{\rm b}$ and  $e_{\rm b}$ refer to the outer planet's mass, semi-major axis and eccentricity, and $m_{\rm A}$, $m_{\rm B}$,  and $a_{\rm AB}$ to the masses and semi-major axis of the inner binary, with $m_{\rm AB}= m_{\rm A}+m_{\rm B}$.\footnote{Eq. \ref{laplace_radius} is obtained by dimensionally solving for the radius at which the gravitational perturbation of the inner stellar binary balances that of the outer companion (more on that in Section \ref{section_dynamics_model}).} Taking masses at face value (Table \ref{table_system_parameters}), we locate $r_{\rm L}$ over a range of planet eccentricity and semi-major axis, and for distinct inner binary semi-major axis within observational uncertainties. The outcome is shown in Figure \ref{fig:radius} with a grey band marking the observed radial extent of the debris disc. 

What clearly stands out in this display is that, within uncertainties on companion semi-major axis and eccentricity, the Laplace radius for this system is smack within the disc, moving in or out with decreasing or increasing planet eccentricity (and the reverse with planet semi-major axis). For mean planet orbital parameters, the location of the Laplace radius divides the disc into two chunks, one dominated by the inner binary and the other by the outer companion. Thus, and going by previous studies of Laplace surface dynamics \citep[e.g.,][]{Tremaine}, we expect a warped structure in the outer part [weak or strong depending on how small or large is the mutual inclination between planet and stellar binary]. Also, with a potentially eccentric planetary companion, one can foresee in-plane eccentricity forcing in the warped segment, and motion that largely respects initial conditions in the inner disc, circular if that is indeed how the disc came to life. Even more interesting is the possibility that for a near circular planetary orbit of a wide enough semi-major axis, the Laplace radius moves close to the disc’s outer edge, leaving the inner binary in full control of the disc, dictating in plane dynamics, all the while significantly hampering warp formation. In sum, we can safely rule out unhampered Kozai-Lidov dynamics over the full range of the disc, with it becoming quenched by the binary over a significant inner range, and for certain planetary orbits, over the whole disc. 
\begin{figure}
\centering
\includegraphics[width=\linewidth]{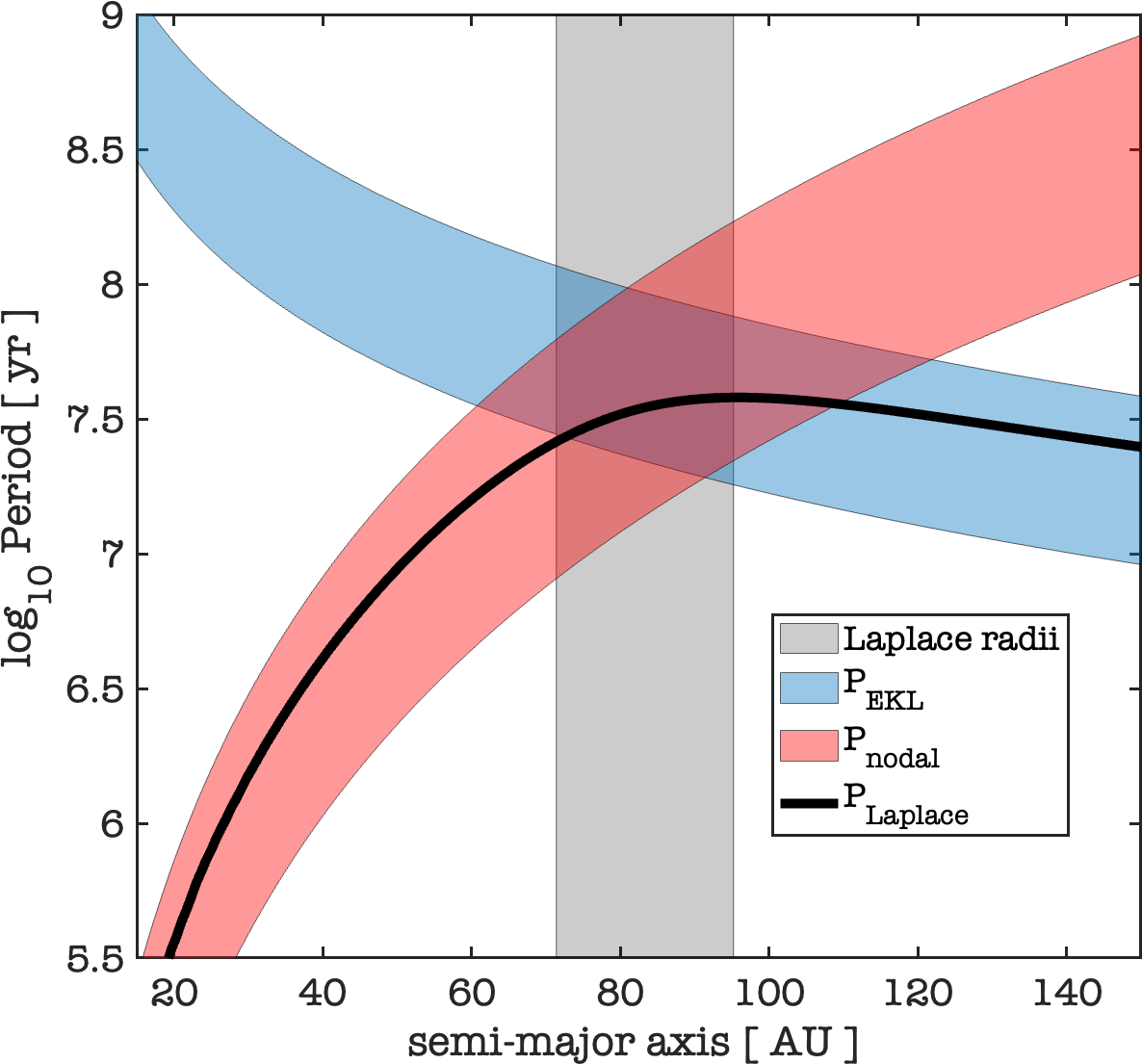}
\caption{Timescales for distinct regimes of debris dynamics. In red, we show the period of binary-induced nodal precession as a function of planetesimal semi-major axis, allowing for a range of semi-major axis for the inner binary and variations of debris orbital elements (Eq. \ref{eq_nodal_precession}). In blue, we show the period of eccentric Kozai-Lidov oscillations driven by the outer planetary companion when allowed to act alone, and for that same range of orbital parameters (Eq. \ref{pEKL}). In black, we plot the period of small amplitude oscillation around stable stationary orbits as maintained by binary and companion acting together. It clearly interpolates between binary and companion controlled timescales. Interestingly this period is comparable to the age of the system in the outer and inner parts of the disc, and few times longer around the Laplace transition. The impact on the disc is ultimately a function of the initial state of the disc, how close or far it is from the system's Laplace surface.}
\label{fig_timescales}
\end{figure}

The Laplace surface in question captures an idealized configuration to which a dissipative particulate disc is expected to settle. However, the binary, and disc along with it, are a mere 13 Myrs old, so it becomes important to assess relevant timescales of planetesimal dynamics, partly to situate the disc temporally vis a vis any steady state, but also to further capture through competing time scales what we emphasised with spatial considerations.  In so doing, it seemed reasonable to isolate the contribution of each perturber before combining them into a continuous whole:
\begin{itemize}
\item We first consider the stellar binary acting alone as it causes the node of a planetesimal to precess about the binary's angular momentum vector with period: $P_{\rm nodal} = {2\pi}/{\Dot{\Omega}}$, where the nodal precession rate,  $\Dot{\Omega}$, is given by \citep{farago2010high}
\begin{equation}\label{eq_nodal_precession}
    \Dot{\Omega} = -\frac{3}{4}n_{\rm AB} \left(\frac{a_{\rm AB}}{a}\right)^{7/2} \frac{m_{\rm A}m_{\rm B}}{m_{\rm AB}^2} \frac{\cos i}{(1-e^2)^2}.
\end{equation}
Here $n_{\rm AB}$ is the mean motion of the inner binary, while $a, e,$ and $i$ refer to the orbital semi-major axis, eccentricity, and inclination of the planetesimal in question. In Figure \ref{fig_timescales}, we display in red the range of binary-induced nodal precession timescales for relevant combinations of debris disc orbital elements (i.e., $a\in[20,150]$ AU; $e\in[0,0.8]$; $i\in[0,80]$) and variations of $n_{\rm AB}$. Evidently, planetesimals closer to the stellar binary precess much faster [$\sim10^6$ yr around 30 AU] than those in the outskirts of the disc [$\sim10^8$ yr around 150 AU]. For particles around the Laplace radius, binary-induced precession timescale is interestingly comparable to the age of the system, $\simeq 10-30$ Myrs. 

\item We then model the binary as a point mass, and consider the forcing of a planetary companion which when sufficiently inclined (and eccentric) can drive planetesimals into Kozai-Lidov cycling with an average period of \citep{antognini2015timescales}:
\begin{equation}\label{pEKL}
    P_{\rm EKL}\approx \frac{256}{15\pi}\sqrt{\frac{10}{\varepsilon_{\rm o}}}t_{\rm sec},
\end{equation}
where the secular timescale $t_{\rm sec}$ is given by 
\begin{equation}
    t_{\rm sec}^{-1}= \frac{m_{\rm b}}{a_{\rm b}^3(1-e_{\rm b}^2)^{3/2}}\sqrt{\frac{Ga^3}{m_{\rm AB}}},
\end{equation}
and the dimensionless coefficient $\varepsilon_{\rm o}$ is defined as
\begin{equation}\label{oct_coeff}
    \varepsilon_{\rm o}=\frac{e_{\rm b}}{1-e_{\rm b}^2}\frac{a}{a_{\rm b}}.
\end{equation}
In Figure \ref{fig_timescales}, we display in blue the range of timescales associated with this regime. As expected, the timescale of K-L cycling is longest where that of binary-induced precession is shortest, and shortest in the outer parts of the disc, where it is slightly longer than the age of the system [$\simeq 2\times 10^7$ yrs around 150 AU]. 

\item Last but not least, we consider the combined field of both inner and outer perturbers, which imposes a typical oscillation frequency on a planetesimal situated in between\footnote{ The frequency in question is that of small amplitude oscillations around the equilibrium orbits maintained over the range of semi-major axis of interest \citep[see Section 3 of ][]{Tremaine}.}. The position-dependent period of this oscillation is marked
by the black curve in Figure \ref{fig_timescales}. Evidently, this curve interpolates, as it should, between binary-induced nodal precession before the Laplace radius (shown with uncertainties in the grey envelope), then Kozai-Lidov cycling  at distances significantly larger than the Laplace radius. A transitional interval intimately combines both perturbers into dynamics that shape the behavior of our disc around the Laplace surface.
\end{itemize}
In sum, Figures \ref{fig:radius} and \ref{fig_timescales} show how the  HD 106906 debris disc is spatio-temporally shaped by the combined gravitational perturbation of both binary and outer perturber, with its relatively young age suggesting a transient architecture which we proceed to map out below.

\section{Towards HD 106906 Debris Dynamics: Secular Dynamical Model and Associated Equilibria }\label{section_dynamics_model}
By now, the reader should find it quite natural, if not unavoidable, to consider the combined gravitational perturbation of binary and planet as one explores the dynamics of the debris disc in HD 106906. The setting is not very different from that of planetary rings, or satellites, in the field of the oblate planet that hosts them and the Sun around which they revolve. Here, we summarize key mathematical machinery leading on one hand to yet another generalized Laplace surface [allowing for an inner stellar binary with an eccentric planetary companion], and on the other setting an economical mathematical model with which to study dynamical evolution of debris (in HD 106906 or any similar such system) in arbitrary detail and for arbitrary durations. 

We introduce an orbital reference frame with the following triad:
\begin{itemize}
    \item ${ \boldsymbol{\hat{n}}}$ in the direction of the planetesimal 's orbital angular momentum.
    \item ${ \boldsymbol{\hat{u}}}$ in the direction of the periapse of its orbit.
    \item ${\boldsymbol{\hat{v}}} = {\boldsymbol{\hat{n}}} \times { \boldsymbol{\hat{u}}}$.
\end{itemize}
Analogous triads are defined in the reference frame of  the inner stellar binary $(\boldsymbol{\hat{n}}_{\rm AB},\boldsymbol{\hat{u}}_{\rm AB},\boldsymbol{\hat{v}}_{\rm AB})$, and for the outer planetary companion $(\boldsymbol{\hat{n}}_{\rm b},\boldsymbol{\hat{u}}_{\rm b},\boldsymbol{\hat{v}}_{\rm b})$.

We are interested in the secular (read orbit averaged) evolution of planetesimals in the debris disc, planetesimals which are assumed far enough from binary and planet to be free of mean motion  or semi-secular resonances \citep[the evection being a prime example of the latter in hierarchical systems;][]{TW98, TS2015}. Such motion is cleanly parameterized with the normalized angular momentum vector, $\boldsymbol{j}=\sqrt{1-e^2}\boldsymbol{\hat{n}}$, and the Lenz vector, $\boldsymbol{e}=e\boldsymbol{\hat{u}}$, with the help of which one avoids coordinate related singularities at zero eccentricity and/or inclination. With these vectorial elements, it is straight forward to recover the Hamiltonian governing the orbit averaged dynamics of disc particles. The procedure, which is briefly outlined in Appendix \ref{Appendix_secular},  generalizes related work on hierarchical triples \citep[e.g.][]{Tremaine,correia2011tidal,hamers2020secular,farhat2021laplace} by allowing for an inner binary and yields the secular Hamiltonian:
\begin{align}\label{basic_Ham}
    H_S= -\frac{Gm_{\rm AB}}{2a} +\frac{Gm_{\rm AB}}{a}\left(\Psi_{\rm AB} + \Psi_{\rm b, quad} +\Psi_{\rm b, oct}\right).  
\end{align}
Here, the inner binary's gravitational field is captured to quadrupolar order through the averaged disturbing function
\begin{align}
    \Psi_{\rm AB}&= \frac{3}{8}\frac{\varepsilon_{\rm AB}}{(1-e^2)^{\frac{5}{2}}}\Bigg[5(\boldsymbol{j}\cdot\boldsymbol{e}_{\rm AB})^2-(1-e_{\rm AB}^2)(\boldsymbol{j}\cdot\boldsymbol{n}_{\rm AB})^2 \\\nonumber
   & +\left(1-e^2\right)\left(\frac{1}{3}-2e_{\rm AB}^2\right)\Bigg],
\end{align}
with dimensionless parameter
\begin{equation}\label{eq_varep_AB}
    \varepsilon_{\rm AB}=\alpha^2 \frac{m_{\rm A}m_{\rm B}}{m_{\rm AB}^2}  \hspace{.2cm} \text{where} \hspace{.2cm} \alpha=\frac{a_{\rm AB}}{a}.
\end{equation}
Note that, although the stellar binary in the HD 106906 system lives on an eccentric orbit \citep[][]{de2019near}, we limit $\Psi_{\rm AB}$ to quadrupolar order because we are considering a stellar twin which has a negligible/null octupolar contribution, depending as it does on the difference in mass between the stellar components \citep[e.g.,][]{correia2016secular}. 

On the other end, we expand the potential of the eccentric inclined planet to octupolar order keeping in mind recent work by \cite{farhat2021laplace} which gives vivid illustrations of how going beyond quadrupoles is essential to Laplace-like dynamics with eccentric outer perturbers. The planet's quadrupolar contribution is given by
\begin{equation}\label{psiquad}
    \Psi_{\rm b, quad}=\frac{3 \varepsilon_{\rm q}}{8}\Big[\frac{1}{3}-2e^2+5e_n^2-j_n^2\Big],
\end{equation}
and its orbit averaged octupolar perturbation by 
\begin{equation}\label{psioct}
    \Psi_{\rm b, oct}= \frac{15}{64}\varepsilon_{\rm q}\varepsilon_{\rm o} \Big[e_u\big(8e^2-1-35e_n^2+5j_n^2\big)+10j_uj_ne_n\Big],
\end{equation}
introducing the dimensionless parameter 
\begin{equation}\label{eq_varep_q}
\varepsilon_{\rm q}=\frac{m_{\rm b}a^3}{m_{\rm AB}a_{\rm b}^3(1-e_{\rm b}^2)^{\frac{3}{2}}},
\end{equation}
together with the previously defined [\eq{oct_coeff}] octupolar coefficient $\varepsilon_{\rm o}$. 
In Eqs. (\ref{psiquad},\ref{psioct}) subscripts affected to $j$ and $e$ reflect bases vectors of the planetary reference frame on which $\boldsymbol j$ and $\boldsymbol e$ are projected respectively. Using the expressions above, we can readily express the Laplace radius of \eq{laplace_radius} in terms of the ratio of inner to outer quadrupolar dimensionless parameters via $r_{\rm L}^5=a^5\varepsilon_{\rm AB}/\varepsilon_{\rm q}$.

It is perhaps worth emphasizing here that one should use a multipolar expansion for the planet's perturbation with caution. Considering the nominal orbital configuration of HD 106906 (refer to Sec.\ref{Section_ELS_106906}), it is safe to assume that the system is sufficiently hierarchical for the multipolar expansion to converge. However, this hierarchy may be compromised with a closer-in and highly eccentric planet, in which case the multipolar series diverges and fails. In this case, one can either resort to exact numerical averaging of the Hamiltonian [as was done in \cite{beust2016orbital,saillenfest2017non,batygin2017dynamical} for the case of Planet 9] , or to the numerical averaging of the spherical harmonics of the planet's potential when modeled as a Gaussian ring \citep{farhat2021laplace}.

With $H_{\rm S}$ in hand, we have a minimal model with which to follow the dynamics of HD 106906-like debris in the secular orbit-averaged regime, one where the orientation and shape of planetesimal orbits are fully described by the evolution of $\boldsymbol e$ and $\boldsymbol j$, all the while their semi-major axis is conserved. Equations of motion are best expressed in the vectorial form \citep{milankovitch1939verwendung,allan1964long}:
\begin{align} \label{Mil_Je} \nonumber
     \frac{d \boldsymbol j}{d\tau}&=-\boldsymbol j\times\nabla_{\boldsymbol j} \Psi-\boldsymbol e\times\nabla_{\boldsymbol e}\Psi, \\
     \frac{d \boldsymbol e}{d\tau}&=-\boldsymbol e\times\nabla_{\boldsymbol j} \Psi-\boldsymbol j\times\nabla_{\boldsymbol e}\Psi,
\end{align}
with $\tau=\sqrt{{Gm_{\rm AB}}/{a^3}}t$, and $\Psi=\Psi_{\rm AB} + \Psi_{\rm b, quad} +\Psi_{\rm b, oct} $. Expanding, we obtain:
\begin{align}\nonumber
    \frac{d\boldsymbol j}{d\tau}&= \frac{3\varepsilon_{\rm AB}}{4(1-e^2)^\frac{5}{2}}\Big[(1-e_{\rm AB}^2)(\boldsymbol j \cdot \boldsymbol{\hat{n}}_{\rm AB})(\boldsymbol j \times \boldsymbol{\hat{n}}_{\rm AB})\\\nonumber
    & -5(\boldsymbol j \cdot\boldsymbol e_{\rm AB})(\boldsymbol j \times\boldsymbol e_{\rm AB})\Big] +
   \frac{3}{4}\varepsilon_{\rm q} j_n\boldsymbol j\times\boldsymbol{\hat{n}}_{\rm b}-\frac{15}{4}\varepsilon_{\rm q} e_n\boldsymbol e\times\boldsymbol{\hat{n}}_{\rm b}\\\nonumber
     &-\frac{75}{64}\varepsilon_{\rm q}\varepsilon_{\rm o}\Bigg\{\Big[2\big(e_uj_n+e_nj_u\big)\boldsymbol j
     +2\big(-7e_ne_u+j_uj_n\big)\boldsymbol e\Big]\times \boldsymbol{\hat{n}}_{\rm b}\\
     &+\Big[2e_nj_n\boldsymbol j+\big(-7e_n^2+j_n^2+\frac{8}{5}e^2 -\frac{1}{5}\big)\boldsymbol e
     \Big]\times\boldsymbol{\hat{u}}_{\rm b}\Bigg\},
     \label{eom_j}
\end{align}
\begin{align}\nonumber
    \frac{d\boldsymbol e}{d\tau}&= \frac{3}{4}\varepsilon_{\rm AB}\Bigg\{ \frac{1}{(1-e^2)^\frac{5}{2}}\Big[(1-e_{\rm AB}^2)(\boldsymbol j \cdot \boldsymbol{\hat{n}}_{\rm AB})(\boldsymbol e \times \boldsymbol{\hat{n}}_{\rm AB})\\\nonumber &-5(\boldsymbol j \cdot\boldsymbol e_{\rm AB})(\boldsymbol e \times\boldsymbol e_{\rm AB})\Big]
    -\frac{1}{2(1-e^2)^\frac{7}{2}}\Big[(1-6e_{\rm AB}^2)(1-e^2)\\\nonumber
   &-5(1-e_{\rm AB}^2)(\boldsymbol j \cdot \boldsymbol{\hat{n}}_{\rm AB})^2 +25(\boldsymbol j \cdot\boldsymbol e_{\rm AB})^2\Big]\boldsymbol j\times \boldsymbol e\Bigg\}\\\nonumber
&+ \frac{3}{4}\varepsilon_{\rm q} j_n\boldsymbol e\times\boldsymbol{\hat{n}}_{\rm b}-\frac{15}{4}\varepsilon_{\rm q} e_n\boldsymbol j\times\boldsymbol{\hat{n}}_{\rm b}+\frac{3}{2}\varepsilon_{\rm q} \boldsymbol j\times\boldsymbol e\\\nonumber
     &-\frac{75}{64}\varepsilon_{\rm q}\varepsilon_{\rm o}\Bigg\{\Big[2\big(e_uj_n+e_nj_u\big)\boldsymbol e
     +2\big(-7e_ue_n+j_uj_n\big)\boldsymbol j\Big]\times \boldsymbol{\hat{n}}_{\rm b}\\
     &+\Big[2e_nj_n\boldsymbol e+\big(-7e_n^2+j_n^2+\frac{8}{5}e^2 -\frac{1}{5}\big)\boldsymbol j
     \Big]\times\boldsymbol{\hat{u}}_{\rm b}
     +\frac{16}{5}e_u\boldsymbol j\times\boldsymbol e\Bigg\}.
     \label{eom_e}
\end{align}
In this perturbative model, the planet’s presence is felt through both the axisymmetric (essentially quadrupolar) terms  with $\varepsilon_{\rm q}$, and the symmetry-breaking terms (here limited to octupolar order) with $\varepsilon_{\rm o}$. The terms in question are identical to those in \citet{farhat2021laplace}, where the reader can further find copious details on their emergence, properties and significance. The contribution of the inner stellar binary is captured through the terms with $\varepsilon_{\rm AB}$. 

\subsection{The Eccentric Laplace Surface of HD 106906 System: A brief interlude}\label{Section_ELS_106906}
\begin{figure}
\centering
\includegraphics[width=\linewidth]{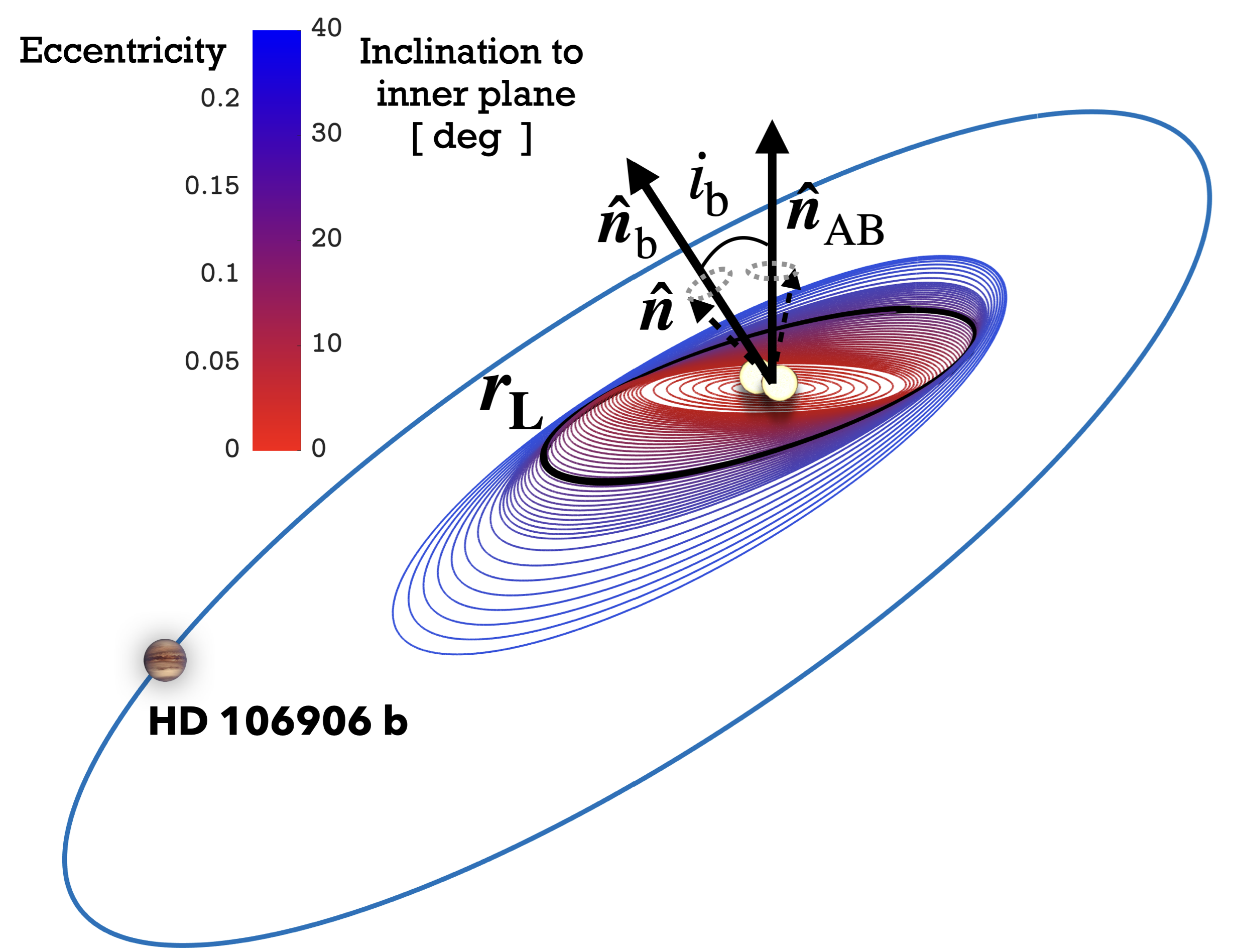} \\
\vspace{.2cm}
\includegraphics[width=.9\linewidth]{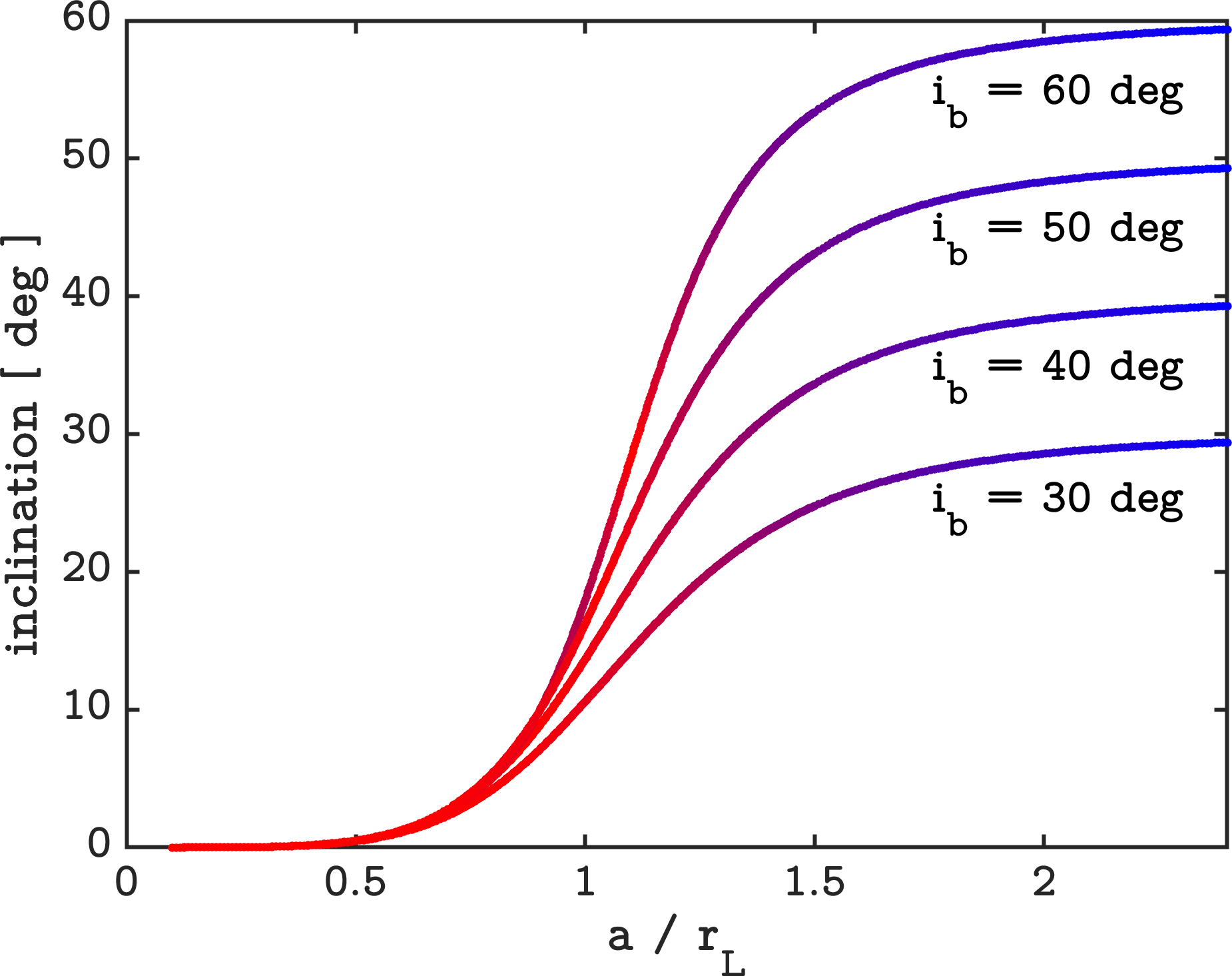} \\

\caption{The Laplace Surface in HD 106906. Planetesimal rings with angular momentum along $\boldsymbol{\hat{n}}$, are torqued to precess about the inner stellar binary with angular momentum along $\boldsymbol{\hat{n}}_{\rm AB}$, and the outer planet with angular momentum along $\boldsymbol{\hat{n}}_{\rm b}$. The mutual inclination between the inner and outer perturbers is defined by the angle $i_{\rm b}$. This misaligned competition promotes precession about the equilibrium Laplace plane. With increasing semi-major axis, equilibrium planes span the warped surface shown the upper panel, with an alignment transition indicated at the Laplace radius $r_{\rm L}$ (Eq. \ref{laplace_radius}). The color codes for planetesimal eccentricity which is pumped as one moves outward on the surface. In the bottom panel, we map planetesimal inclination on the Laplace surface for different values of planetary inclination. The color renders the corresponding eccentricity as in the upper panel. Curves are obtained by solving Eqs. (\ref{equilibrium_1}-\ref{equilibrium_2}) for Laplace equilibria.}
\label{Laplace_sketch}
\end{figure}
With simultaneous perturbations from binary and distant planet, the orbit of a planetesimal, conveniently envisioned as an eccentric and inclined ring, wants to precess about the normal to the stellar binary's plane and the normal to the planetary companion's plane simultaneously, with rates that are weighted by the relative position of the planetesimal. The compromise is that it precesses around the intermediate Laplace plane, the plane of the equilibrium orbit at any given location. The envelope shaped by those planes at any planetesimal position defines the Laplace surface. It was characterized in full generality (instabilities included) for axisymmetric perturbations in \citet{Tremaine}, and later generalized for eccentric perturbations in \citet{farhat2021laplace}. 

Before embarking on debris dynamics which is our main concern, we report briefly on debris statics, mapping out Laplace surfaces in representative HD 106906 systems, the fiducial one included. The purpose is two-fold: 
\begin{itemize}
\item Such orbits represent the skeleton around which debris dynamics is structured; understanding them as a function of planetary companion orbits reflects succinctly the location and extent of spatial warp and in plane distortion of debris distributions; 

\item Classically, the Laplace surface is where dust, ring systems, planetary satellites, are expected to settle under adiabatically slow, non-conservative, processes. With the above discussion of relevant timescales in mind (Figure \ref{fig_timescales}), we expect the inner part of the debris disc to follow the Laplace surface closely, and anything around the Laplace transition and beyond to be in a transient non-adiabatic regime, where excitations can be either more or less significant than indicated by equilibria, and this as a function of disc initial conditions. In either case, the Laplace surface is again a useful reference, with detailed observations (as and when they come in) providing an indication of evolutionary processes that shaped the disc. 

\end{itemize} 

Without serious loss of generality, we limit our discussion to coplanar-coplanar equilibria, in which all the vectors $\boldsymbol{j}, \boldsymbol{e}, \boldsymbol{\hat{n}}_{\rm b}, \boldsymbol{\hat{u}}_{\rm b},$ and $\boldsymbol{\hat{n}}_{\rm AB}$ lie in the same plane. This implies that the nodes of the three players are aligned. The practical reader will agree that coplanar-coplanar equilibria are most relevant here given the fiducial orientation of the planet’s orbit, though nothing prevents further complexification when the situation calls for it! 

Imposing nodal alignment, we recover conditions for stationary debris orbits in the form of two scalar equations in two unknowns: the debris eccentricity $e$, and the debris inclination with respect to the inner plane $i$. For a mutual inclination $i_{\rm b}$ between the stellar binary's plane and the planet's orbit, conditions reduce to:
\begin{align}
\nonumber
    0= &\frac{3}{4}\frac{\varepsilon_{\rm AB}}{(1-e^2)^\frac{5}{2}}\Big[\frac{1}{2}(1-e^2)\sin{2i}(4e_{\rm AB}^2+1)\Big]\\\nonumber
    -&\frac{3}{8}\varepsilon_{\rm q}(1+4e^2)\sin2\phi
\mp\frac{75}{64}\varepsilon_{\rm q}\varepsilon_{\rm o} e\Bigg[-\frac{7}{2}(1+e^2)\cos\phi\sin2\phi\\
+&(2+5e^2)\sin^3\phi+\frac{1}{5}(1-8e^2)\sin\phi\Bigg],
\label{equilibrium_1}
\end{align}

\begin{align}
\nonumber
    0= &\frac{9}{8}\frac{\varepsilon_{\rm AB} e}{(1-e^2)^2}\Big[-\sin^2{i}(4e_{\rm AB}^2+1)+e_{\rm AB}^2+2/3\Big] \\\nonumber
    +&\frac{3}{4}\varepsilon_{\rm q} e\sqrt{1-e^2}(1-4\sin^2\phi)\\\nonumber
\pm&\frac{75}{64}\varepsilon_{\rm q}\varepsilon_{\rm o}(1-e^2)^\frac{1}{2}\Bigg[(3e^2-1)\cos^3\phi\\
+&(\frac{1}{5}-\frac{24}{5}e^2)\cos\phi
+(1+\frac{15}{2}e^2)\sin\phi\sin2\phi\Bigg],
\label{equilibrium_2}
\end{align}
where $\phi=i_{\rm b}-i$.  With nodal alignment, equilibria can be either apsidally aligned or anti-aligned with the outer planet's periapse. We distinguish between them by the sign of the octupole terms, with the upper sign delivering the aligned configuration.

Solving those conditions as a function of planetesimal semi-major axis yields eccentric Laplace surfaces shown in Figure \ref{Laplace_sketch}. Surfaces were computed for various inclinations of the outer companion to the binary plane. They clearly interpolate between an inner circular disc region in near-alignment with binary plane,  with an outer eccentric and inclined region, with mean inclination and eccentricity growing with the planet's inclination. Careful treatment will reveal instabilities in equilibria of a high enough inclination and eccentricity (and bifurcations of host surfaces), the discussion of which will take us too far afield. As indicated above, those surfaces, and the architecture they scaffold, allow economic expression of disc warp and elongation with distance away from the binary. 

The equilibria were here recovered for fixed planet orientation, eccentricity, semi-major axis and varying inclination (Table \ref{table_system_parameters}). They can be further mapped for varying planet eccentricity and semi-major axis, shifting the Laplace transition in or out, fully enclosing the disc in a binary dominated region with little warp and eccentricity, or further exposing it to the onslaught of the planet. It should be evident how this profile and its dynamical implications can be used to rule out planetary configurations otherwise allowed within uncertainties. We prefer to leave the discussion of such variations and their signatures to secular dynamical simulations which we report on next.

\begin{figure*}
\centering
\includegraphics[width=\linewidth]{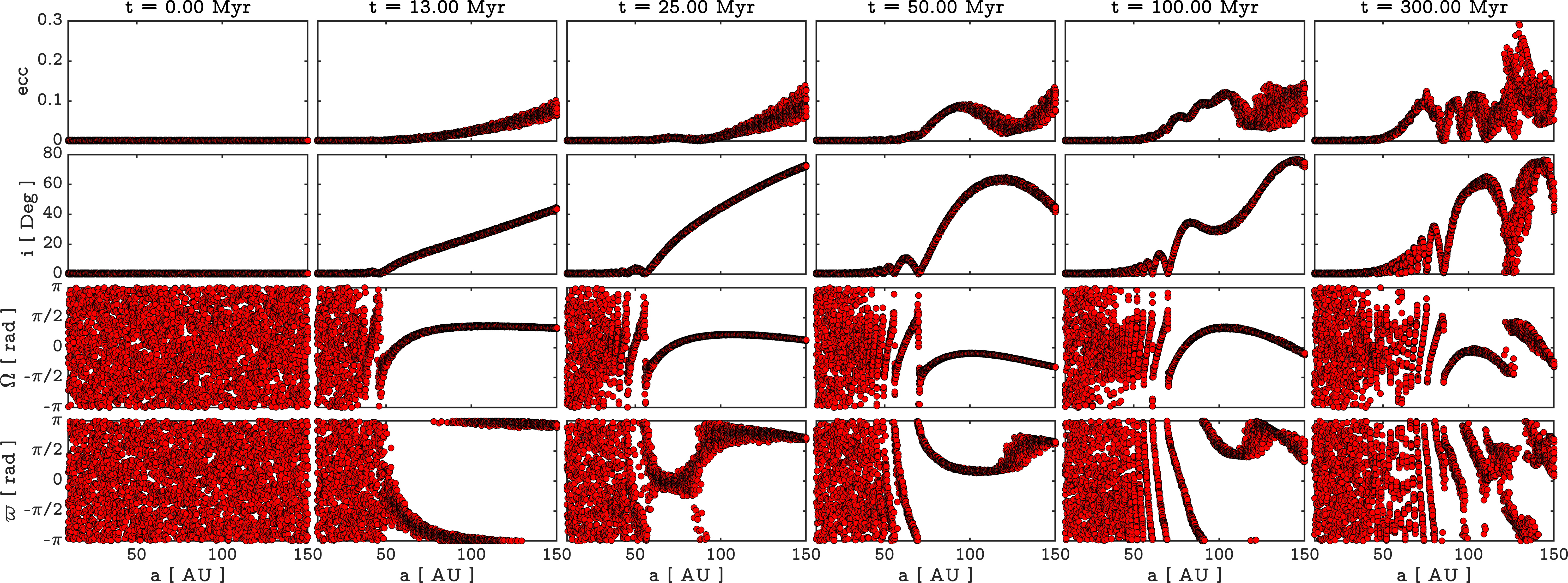}
\caption{Snapshots of the secular orbital evolution obtained via direct numerical integration of the eccentricity, inclination, nodal, and apsidal orientation of distributions of mass-less debris particles perturbed by the stellar binary on the inside and the planetary companion on the outside (Eqs. \ref{eom_j} and \ref{eom_e}). The perturbers' orbits are described in Table \ref{table_system_parameters}.
The disc is initiated with near-circular low-inclination particles with uniform distributions of apsidal and nodal orientations. With the two competing perturbers, and as expected around a Laplace surface, the disc separates into two belts with distinct dynamical regimes: (i) an inner-binary-dominated belt with near-circular, near-coplanar planetesimals characterized by uniform nodal and apsidal orientations; and (ii) an outer planet-dominated belt that undergoes quasi-periodic eccentricity and inclination pumping. Dynamical regimes translate into morphological asymmetries, warps, and spiral arms visualized in Figure \ref{surface_density_maps_binary_on}. The evolution is to be compared with that in Figure \ref{evolution_integration_Bin_OFF} when we mute the effect of the stellar binary. 
}
\label{evolution_integration}

\end{figure*}

\begin{table}
	\centering
	\caption{System Parameters. }
	\label{table_system_parameters}
	\begin{tabular}{lccr} 
		\hline
		Parameter &Unit &Value & Reference \\
		\hline
		Age & Myr & $13\pm 2$ & \cite{pecaut2012revised}\\
		$m_{\rm A}$ & M$_\odot$ & 1.34  &\citet{Beust-2017}\\
		$m_{\rm B}$ & M$_\odot$ & 1.37  &\citet{Beust-2017}\\
		$a_{\rm AB}$ &AU & 0.36-0.58 & \citet{Beust-2017} \\
		$e_{\rm AB}$ & $-$ & $0.66\pm0.002$  &\citet{de2019near}\\
		$m_{\rm b}$ & M$_{\rm J}$ & $11\pm 2$  &\citet{bailey2013hd}\\\vspace{.1cm}
		$a_{\rm b}$ & AU & $850^{+560}_{-260}$  &\citet{Nguyen_2020}\\\vspace{.1cm}
		$e_{\rm b}$ & $-$ & $0.44^{+0.28}_{-0.31}$  &\citet{Nguyen_2020}\\\vspace{.1cm}
		$i_{\rm b}$ & deg & $39^{+20}_{-15}$  &\citet{bryan2021obliquity}\\
		\hline
	\end{tabular}
\end{table}

\section{Dynamics around HD 106906 Laplace surface}
\label{section_dynamics_around_LS}
Having isolated the Laplace surface controlled by inner binary and outer planet (Figure \ref{Laplace_sketch}), our goal now is to study the evolution of a debris disc by numerically integrating the equations of motion (\ref{eom_j}-\ref{eom_e}). We perform a suite of numerical simulations to follow the evolution of the orbital elements of a distribution of $24100$ planetesimals situated between the disc's inner and outer edges, $a_{\rm in}=30$AU and $a_{\rm out}=150$AU respectively. Each planetesimal is parameterized by its secularly constant semi-major axis $a$, and four variable elements: $e,i,\Omega, $ and $\omega$. The equations of motion are solved using an explicit Runge-Kutta (4,5) integrator with adaptive time steps, and constrained by $j^2+e^2=1$ and $\boldsymbol j\cdot\boldsymbol e = 0$. 

We list key ingredients in reported simulations:

\begin{itemize}
\item As the system is sufficiently hierarchical, we ignore the weak mutual perturbation and the consequently ultra-slow precession of inner stellar binary and outer planet  ($\simeq 10^{10}$ yrs)\footnote{In some configurations, that of Planet 9 in particular, the secular evolution of either perturber could be fast enough to be of concern. The matter was discussed at length in \cite{farhat2021laplace} who showed that, in case the secular evolution of a perturber is significantly slower than that of disc particles, the effect is rather minor, with equilibria and dynamics around them responding adiabatically to the slow quasiperiodic evolution of the perturber.}. Thus $(\boldsymbol{\hat{n}}_{\rm AB},\boldsymbol{\hat{e}}_{\rm AB})$ and  $(\boldsymbol{\hat{n}}_{\rm b},\boldsymbol{\hat{e}}_{\rm b})$ are fixed by orbital parameters as given in Table \ref{table_system_parameters}, with the further prescription of $(\Omega_{\rm b}, \omega_{\rm b})= (0,90)$ degrees for the planet. The Laplace radius of this fiducial system is located at $r_{\rm L}=84$ AU;

\item Simulations are agnostic of scenarios for the origin of HD 106906b, some of which might lead to disruption of the debris disc as the planet finds its way to its current location. Instead, the planet is assumed in its observed position for the duration of reported simulations, and unless otherwise stated, with the mean orbital elements of Table \ref{table_system_parameters}; 

\item We ignore any role (gravitational or otherwise) for any gaseous component and/or its dispersal on the dynamics of planetesimals in our simulations, and limit its presence to an item in the discussion section below; 

\item In this first exploration of an already rich dynamical system, we further ignore the effect of debris disc self-gravity on Laplace surface dynamics, and dedicate space in the discussion to assess its potential impact on the dynamics;

\item Simulations do not allow for radiation pressure, the effect of which is included in post-processed calculations reported below; 

\item The debris disc is assumed initially axisymmetric, rather cold, and coplanar with the binary’s orbital planar. Planetesimal orbits are initiated with uniformly distributed eccentricities and inclinations, ranging  between 0 and 0.05, and 0 and 2 degrees, respectively. We consider and report on secular dynamics over a range of semi-major axes which are of relevance to observations, but nothing prevents extending that range in either direction while respecting model assumptions. 

\item In dialogue with published literature, we further perform and report on simulations in which we turn off the quadrupolar perturbation due the inner binary: rather trivial to do numerically, with significant dynamical implications which we comment on in due course. Results which are pertinent to HD 106906 are reviewed first, and are then contrasted with simulations that only account for the binary's monopole in the discussion section. 
\end{itemize}

 With the above assumptions and resulting minimal numerical model, we are able to zero-in on key features pertaining to dynamics around the Laplace surface and potential signatures in the disc’s morphology, current and future. 
Results of simulations can be viewed through evolving distributions of orbital elements and their histograms, then surface density/brightness maps at various stages of evolution. Alternatively, one can view evolution within the system’s phase space, at distinct and telling semi-major axes. We start with orbits and their photometric signature (Section \ref{section_orbital_evo}), then move to phase space structure and the horizons it opens up for future collisional modeling (Section \ref{section_phase_space}). 

\subsection{Orbital evolution and surface density}\label{section_orbital_evo}
Considered together, results presented in Figs.[\ref{evolution_integration},\ref{surface_density_maps_binary_on}] allow the following observations:

\begin{enumerate}

\item {\bf On Orbital Elements} 
\begin{enumerate}

\item {\textbf{Inclinations.}} Currently and far into the future, the stellar binary's quadrupole forces the inner part of the disc to evolve in its orbital plane, all the while making its signature felt into both the amplitude and frequency of oscillations in the outer part. This  clear break around the system's Laplace radius translates into a well defined and dynamically evolving warp between an inner flat belt and an outer thicker distribution evolving around the system’s Laplace surface. The break is unavoidable, and is certainly missed in any modeling that ignores the binary’s non-spherically symmetric contributions as we shall discuss below. 

\item {\textbf{Eccentricities.}} The inclination break around the Laplace radius is further manifested in the evolution of disc eccentricities. The inner binary's quadrupole forces fast precession of inner debris rendering them relatively immune to the planet’s perturbation, further maintaining the initial nearly circular condition of the inner belt, while capping the quasi-periodic eccentricity pumping of the outer part to relatively smaller amplitudes. This is so over the age of the system, and essentially maintained over a few hundred million years, except for a burst in eccentricity which reflects long-term chaotic diffusion in a mixed phase (more on that aspect in Section \ref{section_phase_space}). The effect is familiar from works on the quenching of Kozai-Lidov dynamics by strong enough quadrupolar perturbations, and is here sustained via Laplace surface rigidity, modifying excitation amplitudes and frequencies all the while seeding the frozen orbits of Laplace equilibria. 
\begin{figure*}
\centering
\includegraphics[width=\linewidth]{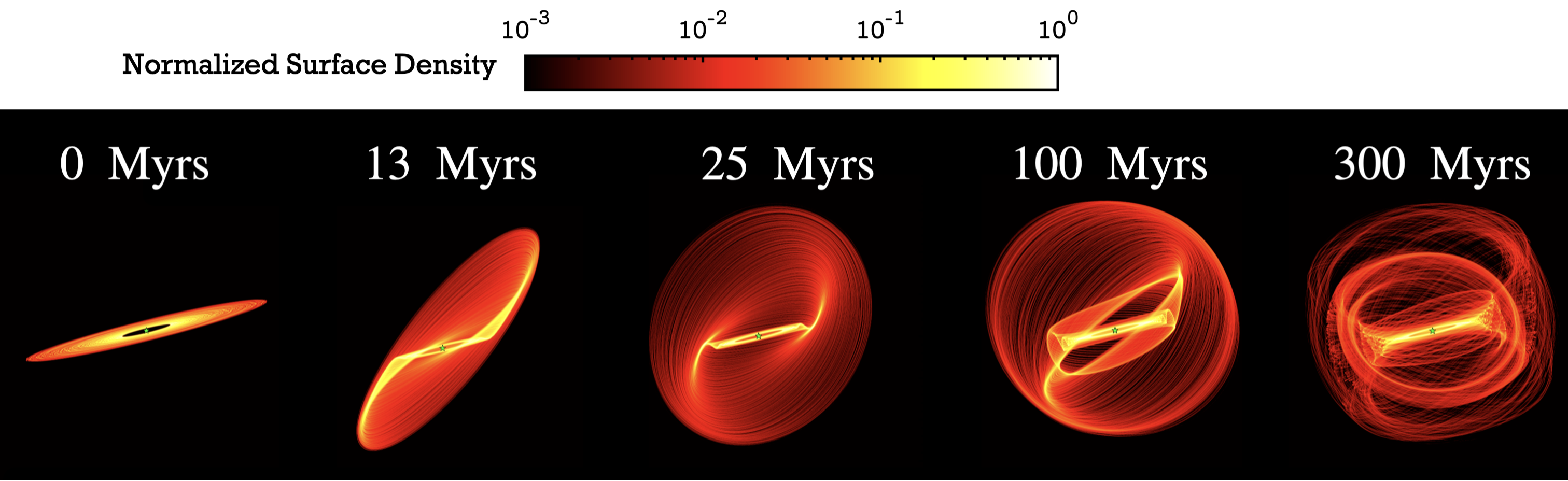}
\caption{Surface density maps showing the time evolution of an HD 106906-like debris disc. The disc is subject to the perturbations of both the inner stellar binary and the outer planet. Maps at a given time and for a set of viewing angles are constructed as described in Appendix \ref{Appendix_Maps} using the distributions of orbital elements resulting from the simulations of Figure \ref{evolution_integration}. The viewing angles are fixed such that: $\theta_{\rm i}=85^\circ$ i.e. the disc is viewed nearly edge-on, $\theta_{\rm l}=13.7^\circ$ as inferred from a position angle of $283.7^\circ$ \citep{kalas2015direct}, and $\theta_{\rm a}=0$ (see Appendix \ref{Appendix_Maps} for further details). Each of these panels have been normalised individually, and the green star represents the center of mass of the inner binary. 
The disc around its current age (13 Myrs) features an inner bar-like structure, courtesy of the inner binary dominated regime, which persists on the long-term evolution of the disc. This structure is terminated the Laplace radius, as the disc warps and transitions to the outer planet dominated regime. Around the Laplace radius also, transient spiral arms are created around the system's current age, and they are twisted with time following the planet's periodic forcing. If allowed to live for 300 Myrs, the separation between the inner and outer regimes becomes more evident: the inner dense bar-like structure is completely separated from the surrounding precessing tori. This figure is to be compared with the morphological evolution of the disc when muting the inner quadrupolar forcing (Figure \ref{surface_brightness_density_13_OFF}), with both scenarios further resolved vertically in the histograms of Figure \ref{Histograms}. An animated version of this morphological evolution is provided within the supplementary material online.}
\label{surface_density_maps_binary_on}
\end{figure*}
\item {\textbf{Nodes and Apses.}} By 13 Myrs, one notes how nodes of planetesimal orbits develop a gradual twist with increasing inclination: a telltale feature of the unfolding warp of classical Laplace dynamics. By then, it is also evident that the growth of eccentricities in the warped outer part of the disc is accompanied by the sculpting of the longitude of the apse into one dimensional manifolds which are suggestive of a developing  two-armed spiral. This is while a nearly axisymmetric coplanar structure is maintained within the Laplace radius. Further evolution maintains a uniform distribution of the apses in the inner belt, all the while the spiral of the outer part twists, winds and folds into a complex triaxial formation. 

 \end{enumerate}

\item{\textbf{On Morphology}} 
\bigskip

We now examine how the dynamics described above [and which, in one form or the other, is controlled by the Laplace surface] manifests through the morphology of the disc in our fiducial configuration. To do this, we construct in Figure \ref{surface_density_maps_binary_on} maps of surface  density using the orbital element distribution of planetesimals portrayed in Figure \ref{evolution_integration}. The procedure which we follow extends what is presented in \citet{Sefilian21} to account for planetesimal inclination [refer to Appendix \ref{Appendix_Maps} for further details]. By 13 Myrs, the system’s current age, the combined gravitational action of binary and outer planet enforces a clear cut separation between an inner bar-like, binary-dominated, structure, and an equally clear warp which is seeded around the system's Laplace radius. The bar-like structure captures in projection the near circular axisymmetric belt sheltered within the binary's influence. The warp reflects the almost linear growth in inclination, with disc particles responding to both binary and companion torque as they drift vertically and precess around the local Laplace plane orientation. The warp further features a spiral arm which reflects eccentricity excitation together with the self-organisation of apses discussed above. The long-term dynamical evolution further exemplifies the morphological features of Laplace dynamics. The inner quadrupole’s presence maintains the bar-like structure, which further supports precessing toroidal structures in the outer parts. 
\end{enumerate}

In sum, the binary imposes a definite break in the disc with clear observational signatures of relevance to ongoing and future campaigns: a distinctive structure in the vertical distribution, with a cold component on the inside giving way to a hot, evidently warped component on the outside; near circular orbits on the inside, with spiral arms seeded on the edge of warps on the outside, the former projecting into a boxy surface density extending till the Laplace radius on the sky, and the later delineating two scythe blades pointing away and towards the observer in density maps. We have much more to say on those observations and variations thereupon, variations which we relegate to a hefty discussion following yet another and, to our mind, essential perspective on the debris disc in its phase space. 

\subsection{Phase-space evolution of a cold disc: look for the separatrix!}\label{section_phase_space}

\begin{figure*}
\centering
\includegraphics[width=\linewidth]{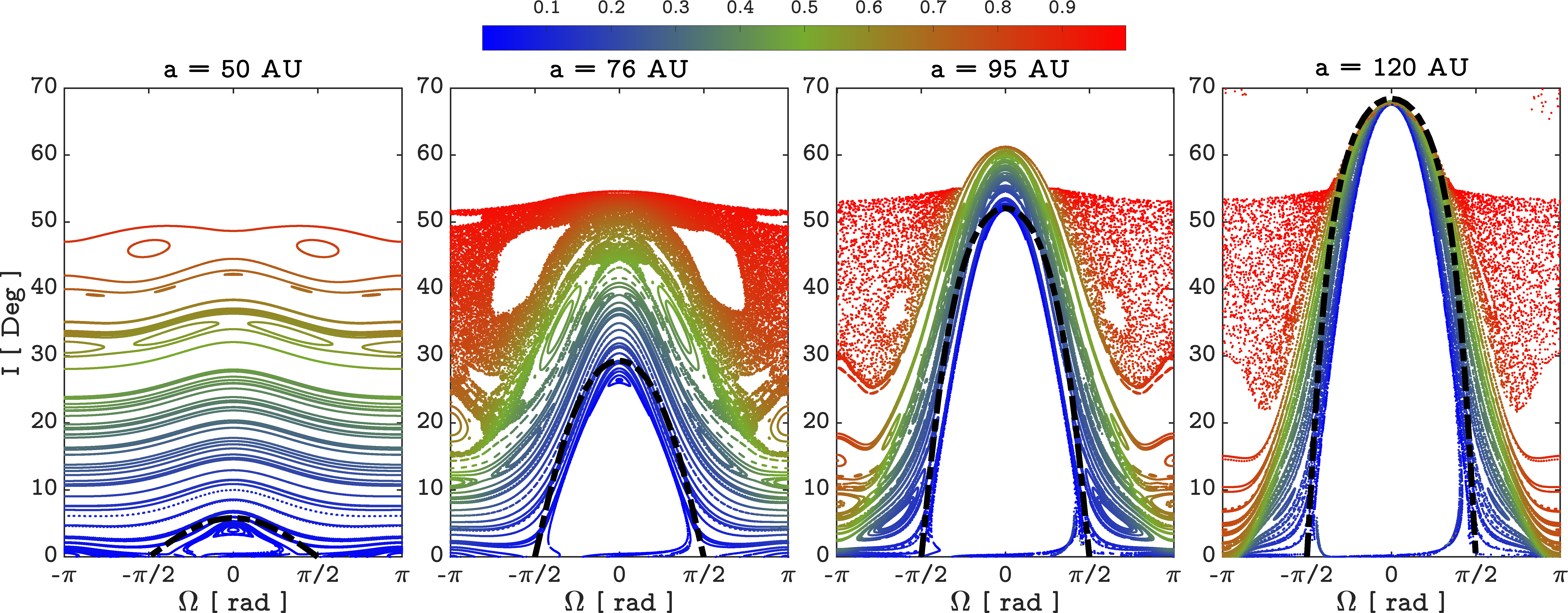}
 \caption{Poincar\'{e} sections for debris particles at different semi-major axis values, driven from the inside by the stellar binary and from the outside by a planetary companion. Initial conditions are selected on an energy manifold that corresponds to low eccentricity-low inclination orbits.  Trajectories are sectioned in $(\Omega,I)$-plane for $\omega=\pi/2$. Crossings are collected when $\dot{\omega} >0$. Color coding corresponds to eccentricity. Of significance to us in these maps are the separatrices joining the nodally aligned and anti-aligned equilibria. If the disc is initiated as circular and flat in the binary's plane, the debris particles at any semi-major axis are destined to live on the separatrix, following its inclination and, to a lesser degree, eccentricity excursions.  We provide in \eq{separatrix_equation} a closed form solution to this structure, and at each semi-major axis, we plot in black the associated trajectory crossings.}
\label{Poincare_sections}
\end{figure*}

To investigate the long-term dynamics of debris particles in phase space, we construct Poincar\'{e} sections on the $(\Omega, i)-$plane, defined by $\omega=90$ degrees, i.e. for debris apsidally aligned with the planet. We restrict plane crossings by imposing the condition of $\dot{\omega}>0$, and we follow the orbits evolving according to Eqs. (\ref{eom_j}-\ref{eom_e}), starting with initial conditions prescribed by an energy hyper-surface. Each surface of section is parametrized by the secularly conserved semi-major axis. We present in Figure \ref{Poincare_sections} samples of sections over a range of debris semi-majors axes which is representative of the disc studied thus far. We further color-code for the eccentricity at section crossings. For each semi-major axis, we construct the section at an energy corresponding to an initially circular non-inclined debris orbit. 

In Figure \ref{Poincare_sections}, we start with a section at $a=50$ AU, focusing on debris which are within the inner quadrupole dominated regime. Debris dynamics on this section is fairly regular, with circulating trajectories covering almost  all of the section space and conserving their vertical angular momentum. A libration island appears around low inclinations and $\Omega=0$. Noting that the section's energy is not the energy of the Laplace equilibrium, this identified relative equilibrium is not exactly the Laplace equilibrium, but can be mapped into it adiabatically. Other relative equilibria appear at low inclinations and $\Omega=\pm\pi$, and a clear separatrix emerges in between. Debris initiated in the plane of the inner binary with near-circular orbits (color coded with blue) would either librate within the low inclination equilibria, or would live on the separatrix. Those living on the separatrix would undergo inclination excursions of a few degrees ($\sim5^\circ)$. 

Moving outwards in the disc, we display in the second panel from the left a section at $a=76 $ AU.  As we are closer to the Laplace radius, the interplay between the inner and outer perturbers is now manifest. The previously regular space of circulating trajectories at higher inclinations is now broken up: a chaotic zone emerges at high inclination with chains of higher order resonances embedded within it. Notable is the persistence of libration islands around nodal alignment and anti-alignment ($\Omega= 0, \pm\pi)$. The libration island by $\Omega=0$ is now around $i\approx 10$ degrees, while those around $\Omega=\pm\pi$ remain at low inclinations. Consequently, debris which is initially in the plane of the inner binary  can now reach $30$ degrees. Increasing the semi-major axis further, we enter the outer planet's dominated regime. Phase space is now encroached upon by area-filling chaotic zones then high order resonant chains around a prominent island of quasidperiodic motion. Our initially cold disc particles (i.e. on near-circular-coplanar orbits) are seen to circulate on the phase-space trajectory which bounds this regular region, a separatrix of sorts ferrying disc particles with increasing mean inclination as we increase $a$, all the while maintaining low to moderate eccentricity. 

From what preceded, it is evident that a one parameter family of phase-space trajectories is critical to the long-term dynamics of debris discs which are initially on near-circular orbits and co-planar with the inner binary. With such initial configurations and their long-term evolution in mind, it would seem reasonable to seek an approximate expression for this family. This is what we proceed to do next as we associate members of this family, at any given semi-major axis $a$, with the "separatrix" of an intimately related integrable dynamical system.  

\begin{figure*}
\centering
\includegraphics[width=\linewidth]{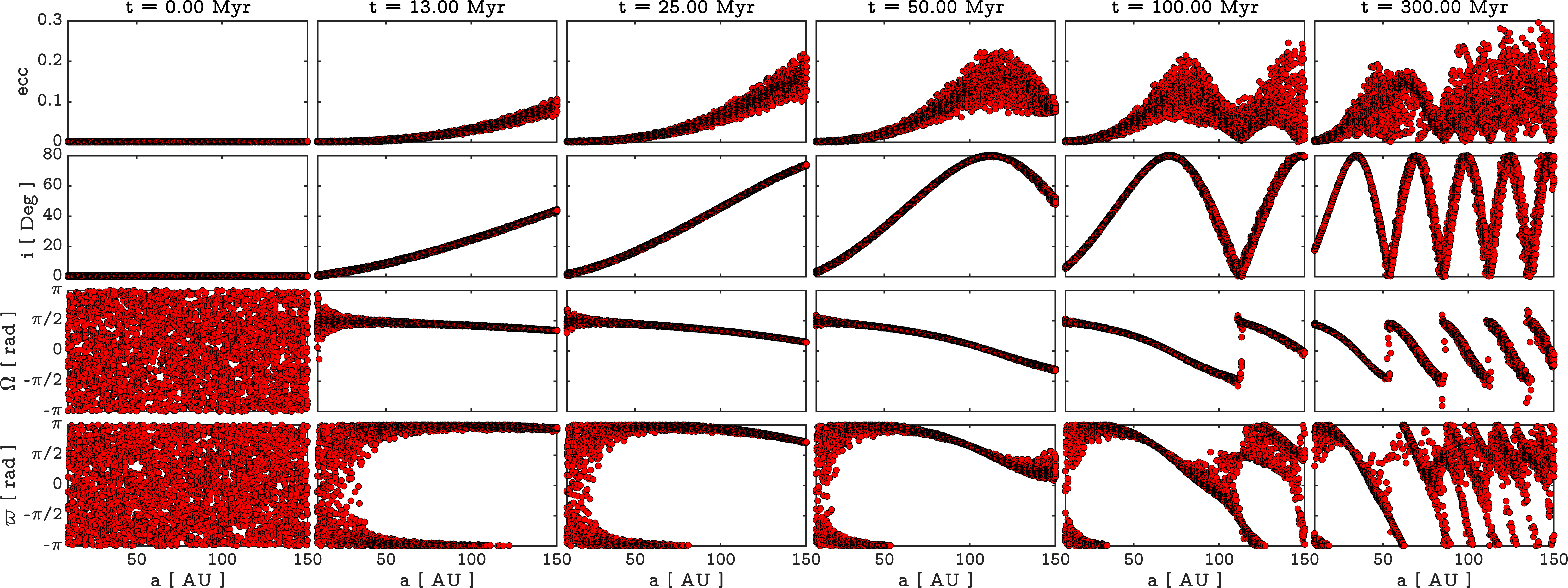}
\caption{Similar to Figure \ref{evolution_integration}, but with the outer planet as the sole perturber to the disc. With the muted effect of the inner stellar binary, the whole disc, inside to outside, is dominated by the torque of the planet: debris particles undergo amplified eccentricity oscillations compared to Figure \ref{evolution_integration}, and are subject to periodic excursions in inclination of amplitude $2i_{\rm b}$.}
\label{evolution_integration_Bin_OFF}
\end{figure*}

\citet{Tremaine} took a first, and already challenging, step towards analytically identifying Laplace equilibria and nonlinear dynamics around them in the limit of circular orbits under quadrupolar perturbations. We can safely work within this limit given how debris orbits maintain relatively low eccentricity in the course of their evolution. As shown in Appendix B of  \citet{Tremaine}, particle dynamics (now fully described by the evolution of the angular momentum vector $\boldsymbol{j}$) can be mapped into the dynamics of an integrable free-rigid body with a suitably defined inertia tensor. Adopting this methodology, and taking $\boldsymbol{\hat{n}}_{\rm AB}=(1,0,0)$, and $\boldsymbol{\hat{n}}_{\rm b}=(\cos i_{\rm b},\sin i_{\rm b},0)$, \eq{eom_j} is rewritten as
\begin{equation}
    \frac{d\boldsymbol{j}}{d\tau} = \frac{3}{4} \boldsymbol{j}\times \boldsymbol{T}\cdot\boldsymbol{j},
\end{equation}
where the symmetric inertia tensor $\boldsymbol{T}$ is given by
\begin{equation}
\boldsymbol{T}=\begin{pmatrix}
\varepsilon_{\rm q}\cos^2 i_{\rm b}+\varepsilon_{\rm AB} & \varepsilon_{\rm q}\cos i_{\rm b}\sin i_{\rm b} & 0\\
\varepsilon_{\rm q}\cos i_{\rm b}\sin i_{\rm b}  &\varepsilon_{\rm q}\sin^2 i_{\rm b} & 0 \\
0&0&0
\end{pmatrix}.
\end{equation}
Denoting by $\boldsymbol{R}$ the orthogonal transformation that diagonalizes $\boldsymbol{T}$, we rotate to the principal-axis coordinate system via $\boldsymbol{j}=\boldsymbol{R}\cdot\boldsymbol{J}$. Under this transformation, the reduced form of  the Hamiltonian in \eq{basic_Ham} is written as \citep{Tremaine}:
\begin{equation}
    \hat{H} = -\frac{1}{2}\boldsymbol{j}^{\rm  T}\cdot\boldsymbol{T}\cdot\boldsymbol{j} = -\frac{1}{2}\sum_{k=1}^{3}t_kJ_k^2 = -\frac{1}{2}\left(t_2J_2^2+t_3J_3^2\right), 
\end{equation}
with eigenvalues $t_1=0<t_2<t_3$ for $i_{\rm b}>0.$ This Hamiltonian generates the (Lie-Poisson) dynamics of the transformed angular momentum $\boldsymbol{J}$ via
\begin{equation}\label{hamiltonian_top}
    \frac{d\boldsymbol{J}}{d\tau}= -\boldsymbol{J}\times\nabla_{\boldsymbol{J}}\hat{H},
\end{equation}
with two integrals of motion: $\hat{H}$ itself, and the magnitude of the angular momentum $J^2$. Orbital trajectories and equilibria can now be identified as the intersection between the angular momentum sphere and the energy surface, an elliptical cylinder. From \eq{hamiltonian_top}, the latter is bounded by the limits $\hat{H}_{\rm min}= -t_3/2$ and  $\hat{H}_{\rm max}= 0$. Between those limits, at the critical energy $\hat{H}_{\rm c}=-t_2/2$, the cylinder is tangent to the sphere, and their intersection defines the sought after separatrix. To obtain the orbits on this separatrix we eliminate $J_3^2$ between the angular momentum and energy integrals to obtain:
\begin{equation}\label{separatrix_equation}
    (t_3-t_2)J_2^2 + t_3J_1^2 = \text{const} = t_3 - t_2. 
\end{equation}
We solve for orbits with angular momentum $\boldsymbol{J}$ on this separatrix, then we project back to orbits with $\boldsymbol{j}$ using the inverse transformation. Solutions yield the phase curves shown in black in Figure \ref{Poincare_sections}. Though obtained under the assumption of circular orbits under dual quadrupolar forcing, those curves are in  remarkable agreement with the phase-space trajectory obtained by integrating the full equations of motion. Specifically, the amplitude of inclination excursions is matched between the two methods over the semi-major axis range of interest.  

\begin{figure*}
\centering
\includegraphics[width=\linewidth]{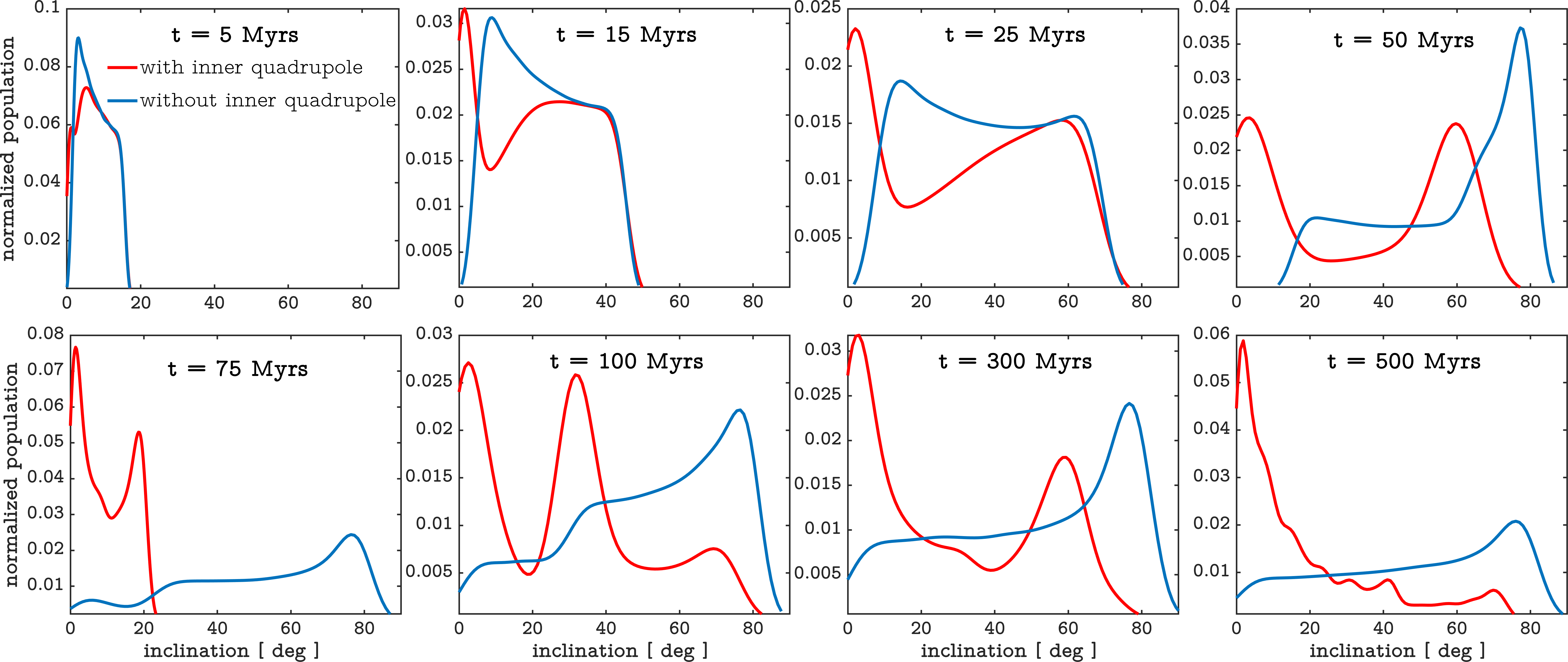}
\caption{Inclination distributions of the disc particles derived from our direct orbital integrations (Figure \ref{evolution_integration}) and sliced at different times. The two kernel-smoothed curves compare disc vertical  distributions in the presence and absence of the inner binary forcing, using the surface density distribution described in Appendix \ref{Appendix_Maps}. The clear discrepancy between the two distributions in each panel signifies taking into account the stellar binary as an inner perturber rather than a single effective central star. Namely, the disc under the two forcings exhibits a bimodal distribution in inclination, with a robust peak around low inclinations, and a fluid peak that shifts between 20 and 70 degrees as time evolves. This is evidently not the case when the inner binary is switched off: the bulk of the distribution smoothly transitions from around low inclinations at $t=0$ Myrs to much higher inclinations as time evolves. }
\label{Histograms}
\end{figure*}

Phase space structure succinctly captures disc observables as it sets the stage for long-term collisional evolution of the disc, whether for the cold initial disc we consider here or for potentially stirred initially distributions that one may wish to envisage. The separatrix recovered analytically here controls collisionless cold disc dynamics and will prove essential for the study of long-term collisional evolution of such discs. Such dynamics will then play out over the richer neighboring structures including the potential for chaotic diffusion to higher eccentricities and/or inclinations. An initially hot disc will start out distributed over phase-space with collisional dynamics reflecting kinetics over distinct dynamical regimes, then diffusion between them.

\section{Variations and Implications: A Discussion}

Here, we revisit modeling assumptions and explore physical scenarios through which we put the results above in dialogue with previous work, assess implications for disc observables (asymmetries and vertical structure in particular), improve constraints on the planet's orbit, and highlight the broad applicability of Laplace surface dynamics.

\subsection{Binary off}
\begin{figure*}
\centering
\includegraphics[width=\linewidth]{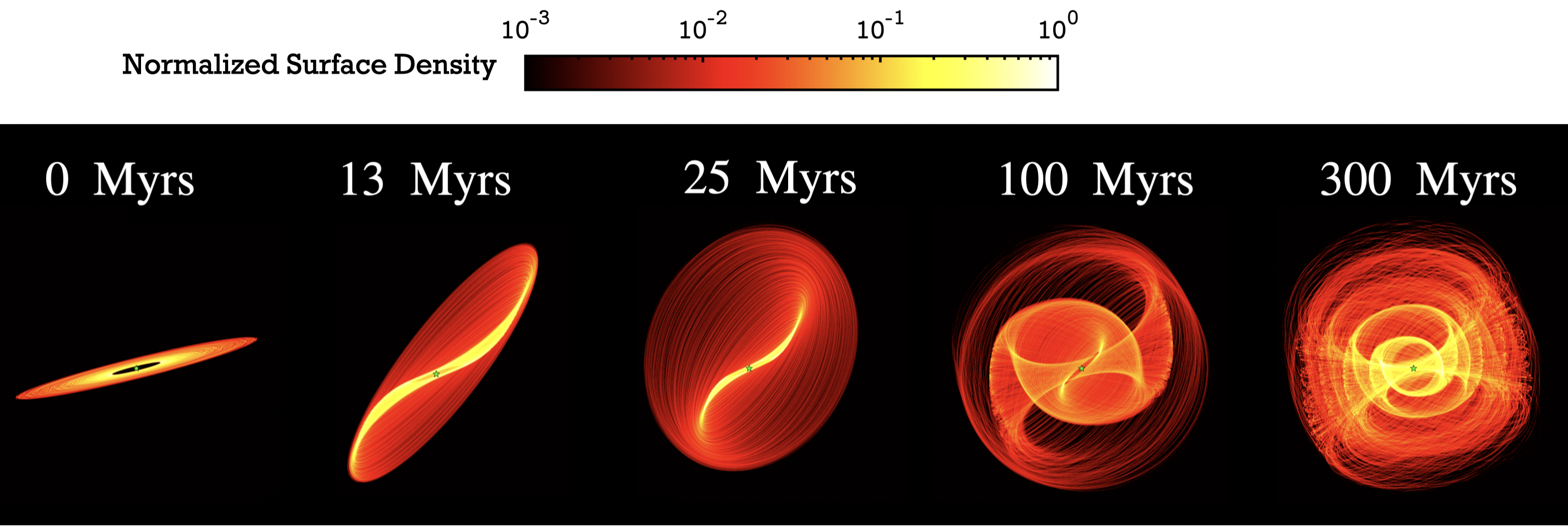}
\caption{Similar to Figure \ref{surface_density_maps_binary_on}, but muting the effect of the stellar binary. The maps are thus constructed from the orbital elements distributions shown  in Figure \ref{evolution_integration_Bin_OFF}. Around the system's age, the whole disc is subject to inclination pumping, leaving little room for maintaining a cold population, and extending the warp onto closer-in radii. Allowing the system to evolve longer, it enters into a steady state where the disc uniformizes in orientations around clear x-shaped and spiral overdensities. An animated version of this morphological evolution is provided within the supplementary material online.}
\label{surface_brightness_density_13_OFF}
\end{figure*}

Neglecting the binary's quadrupole, as was mistakenly done in earlier studies, exposes the disc to the planet's onslaught with effects manifest over the full extent of the disc (extent which is still hierarchical enough to fall under secular dynamical considerations):

\begin{enumerate}

\item With the binary reduced to its monopole, we end up with forced eccentricities that are twice as large, and over the whole expanse of the disc in question. This is evident in Fig.\ref{evolution_integration_Bin_OFF} which further shows how the planet acting alone forces planetesimals over the full range of the disc to oscillate about its orbital plane. Inclination evolution can be well approximated with $i=2i_{\rm b}|\sin\left(\nu t/2\right)|$ \citep{murray1999solar}. Here $\nu$, the secular oscillation frequency, is dependent on the simulated disc lifetime \citep{dawson2011misalignment}. By 13 Myrs, the age of HD 106906, all planetesimals have undergone their first oscillation cycle, before reaching a peak inclination amplitude equal to $2 i_{\rm b}$. By 300 Myrs, all planetesimals have completed multiple cycles, with those closer to the planet oscillating at a higher frequency, creating a vertically thick inclined disc. The presence of the inner perturbation modifies this picture in ways which will have profound implications for observations of HD 106906, as they raise questions for earlier approaches to the modeling of this system.

\item To further clarify distinctive observable features which are maintained by the tight binary in this relatively distant debris distribution, we revisit our simulations via inclination histograms. In each panel of Figure \ref{Histograms}, we show the disc’s inclination distribution,  in the presence and absence of the stellar binary's quadrupole, adopting the surface density defined in \eq{eq:Sigmad_PL}. The discrepancy between the two settings emerges very clearly beyond 5 Myrs of the disc's evolution time. In the presence of the inner binary, the bimodal distribution appears to be a robust feature which is sustained up to $\sim$500 Myrs. The cold -- low inclination -- population persists at all in all phases of the simulation with a mean around $\sim$ 5 degrees. In contrast, the hot population has a mean which shift back and forth in time, ranging between 20 and 70 degrees. In the absence of the inner binary's quadrupole, and as reported in the previous section, the whole disc is subject to unhampered  inclination perturbations by the planet, and it is only a matter of time before the initially flat disc loses its cold population. As seen in Figure \ref{Histograms}, the density peak around small inclinations smoothly transitions in time towards high inclinations and persists there, leaving no remarkable features on the cold end of the distribution. 

\item Morphologically, when the inner quadrupole is switched off, the warp is much stronger and extends to the inner regions by the age of the system (Figure \ref{surface_brightness_density_13_OFF}).  On the long-term, a winding spiral shapes the disc's evolution into a triaxial distribution which leaves a distinctive x-shaped overdensity in projection. While an inner bar is typical of Laplace surface dynamics with competing inner and outer perturbers, the x-shape in the brightness profile of an edge-on disc is telling of the long-term impact of an inclined massive outer planet \citep{pearce2014dynamical}. 

\end{enumerate}

\begin{figure*}
\centering
\includegraphics[width=\linewidth]{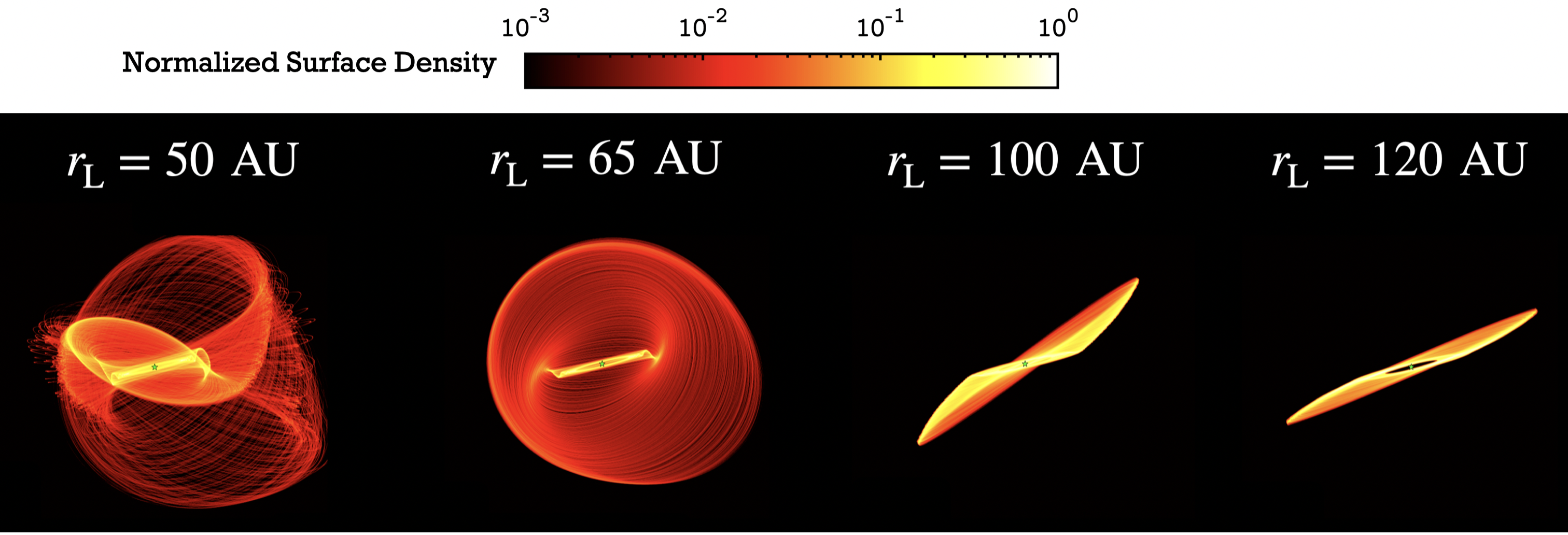}
\caption{ Surface density maps of the disc at its current age as a function of the system's Laplace radius. We control the latter by varying the orbital parameters of the outer planet within the uncertainty intervals reported in Table \ref{table_system_parameters}. These intervals allow for values of $r_{\rm L}$ very close to the inner and outer edges of the disc (see also Figure \ref{fig:radius}), thus placing the disc completely under the dominance of either the stellar binary or the planet. With $r_{\rm L}\rightarrow a_{\rm in}$, we end up with a planet-dominated disc which features warps and asymmetries, and for $r_{\rm L} \rightarrow a_{\rm out}$ we have a fairly rigid, binary-dominated disc which maintains its initial axisymmetric structure.} 
\label{surface_density_varying_rL}
\end{figure*}

\begin{figure}
\centering
\includegraphics[width=\linewidth]{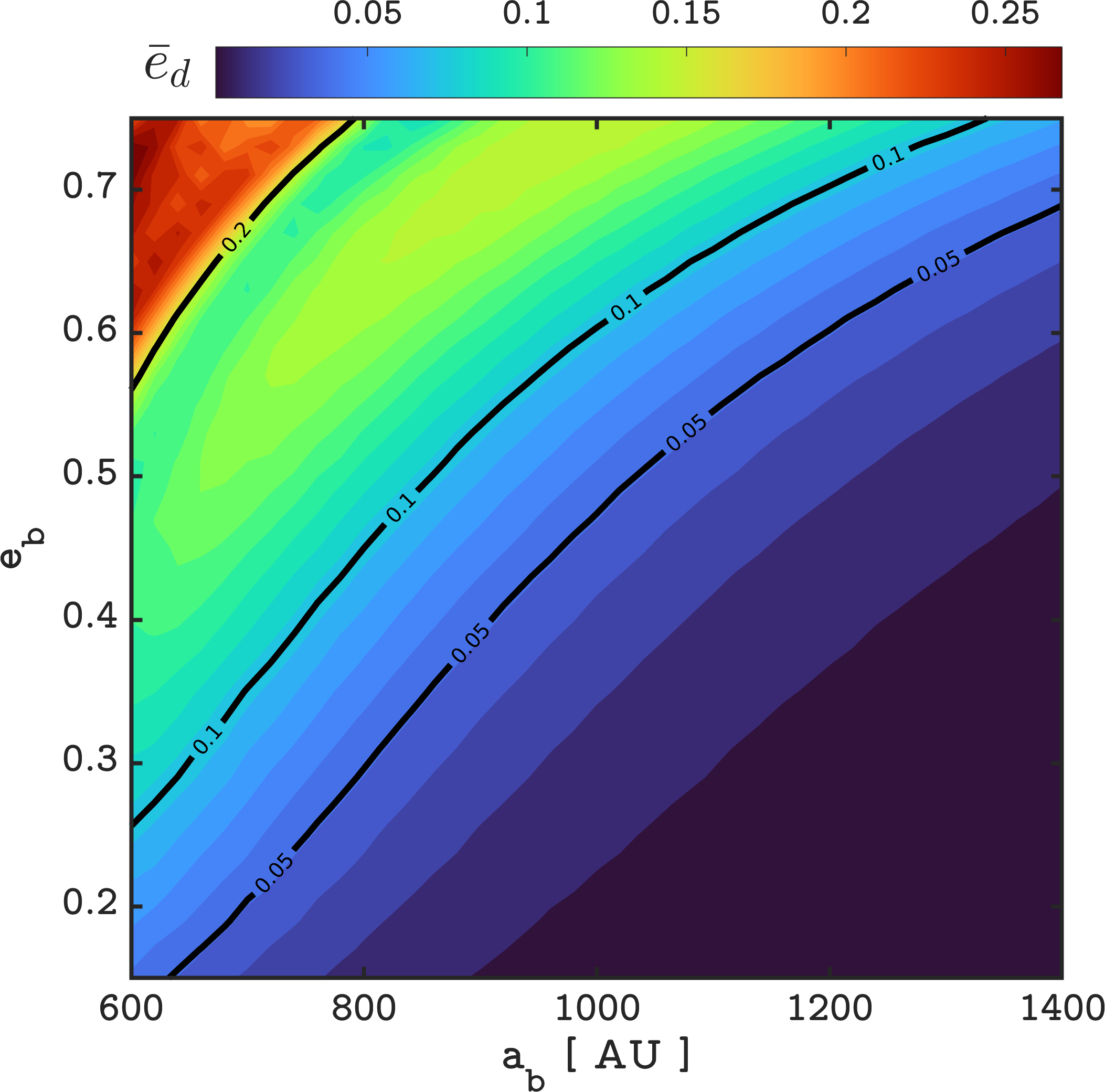}
\caption{Constraining the orbital parameter space of HD 106906 b with disc eccentricity estimates. We contour the surface of disc eccentricity $\bar{e}_{\rm d}$ computed for a range of values of the planetary semi-major axis $a_{\rm b}$ and eccentricity $e_{\rm b}$.  The latter ranges are obtained from the uncertainty intervals in \citet{Nguyen_2020}. The modelled disc is subject here to pure secular gravitational forcing from the inner binary and the outer planet as described in our Laplace machinery. Thus, eccentricity estimates obtained here can, in principle,  be matched with those recovered from ALMA observations.}
\label{disc_ecc_grav}
\end{figure}

\subsection{Disc asymmetries, radiation pressure, and constraints on HD 106906 b orbit }\label{variations_planet_rL}
The HD 106906 debris disc was observed in scattered light to feature asymmetry in brightness, with the southeast extension appearing brighter than the northwest extension in near-infrared intensity \citep{kalas2015direct, Lagrange-2016}. \citet{kalas2015direct} further suggested that the faint west side extension  creates a ``needle"-like structure. Observed asymmetries suggest that the disc is eccentric, leading to pericenter glow which explains brightness bias, together with a faint tail extending towards the apocenter side of the disc \citep{wyatt1999observations}. More recent polarimetric data analysis over multiple bands confirmed the asymmetry and suggested that the disc can be fitted, in surface brightness and structure, to an eccentric ring of eccentricity $\bar{e}_{\rm d}\geq0.16$ \citep{crotts2021deep}. In contrast with this structure in the $\mu$m, no obvious asymmetry was detected using ALMA imaging in the mm \citep[][]{kral2020survey}, with expectations for higher resolution images to resolve the matter once and for all\footnote{While this work was in the final stages of preparation, \cite{meredith_106906} reported on their ALMA observations of the disc at a resolution of 0\farcs38 [39 AU]. They found that  the observed structure can be fitted to an axisymmetric, radially broad disc, without any statistical evidence of an asymmetry or eccentricity, consistent with the analysis of \cite{kral2020survey}.}.

In our study thus far,  we took the best-fit planetary parameters listed in Table \ref{table_system_parameters} for granted, ignoring reported uncertainties \citep{Nguyen_2020}. However, within the wide uncertainty intervals in $a_{\rm b}$ and $e_{\rm b}$, the Laplace radius of the system can take any value between the reported disc's inner and outer edges. In Figure \ref{surface_density_varying_rL}, we simulate the disc for various values of $r_{\rm L}$ ranging between 50 AU and 120 AU, by varying the planetary parameters within the allowed range (Table \ref{table_system_parameters}), and we regenerate surface density maps at the system's current age. Pushing $r_{\rm L}$ to 50 AU, which would result from $(a_{\rm b}, e_{\rm b})=(600$ AU, 0.72),  a smaller chunk of the disc remains controlled by the inner binary, and the larger part becomes planet-dominated. With the planet's percienter closer in, the timescale of the secular forcing by the planet also becomes shorter, thus the disc structure observed at 100 Myrs in Figure \ref{surface_density_maps_binary_on} can now be seen at 13 Myrs. Pushing $r_{\rm L}$ closer to the outer edge of the disc, which can be attained with a planet of $(a_{\rm b}, e_{\rm b})=(1150$ AU, 0.25), allows the binary to overshadow the planet's effect throughout the entire disc, forcing a near flat and circular disc. Thus, and going by \cite{kral2020survey} who see no statistically significant indication of disc eccentricity, it would seem that the system's Laplace radius sits at or beyond the edge of the debris disc, hence favoring a rounder and more distant planet \footnote{\cite{meredith_106906} went through this same exercise and reached a similar conclusion on the planet's orbit based on their ALMA observations of the disc.}.

In sum, we have shown how questions about the existence or not of warps and/or asymmetries in the disc reduce to an estimate of a single powerful parameter, the system's Laplace radius of \eq{laplace_radius}, which in turn 
helped constrain the planet's orbit with a broad range of uncertainties. Evidently, we have here a rather powerful criterion which deserves further refinements in dialog with emerging and future observations of this and similar such discs. 

For now, and with the indications we have available to us, we undertook a systematic exploration of the asymmetry in our simulated disc with the aim of correlating global disc eccentricity with the parameters (mainly orbital) of the planetary companion. We approached the question from two different perspectives. First, we examined disc eccentricity in a suite of secular dynamical simulations exploring a range in $(a_b, e_b)$. We did so in the hope of constraining the planet's orbital parameters by confronting simulated disc eccentricity with ALMA observations of disc geometry. Then, we allowed for radiation pressure (in the addition to the purely gravitational interactions in our model) and undertook the same parametric exploration to compare with scattered light imaging \citep{kalas2015direct, crotts2021deep}. 

\begin{figure}
\centering
\includegraphics[width=\linewidth]{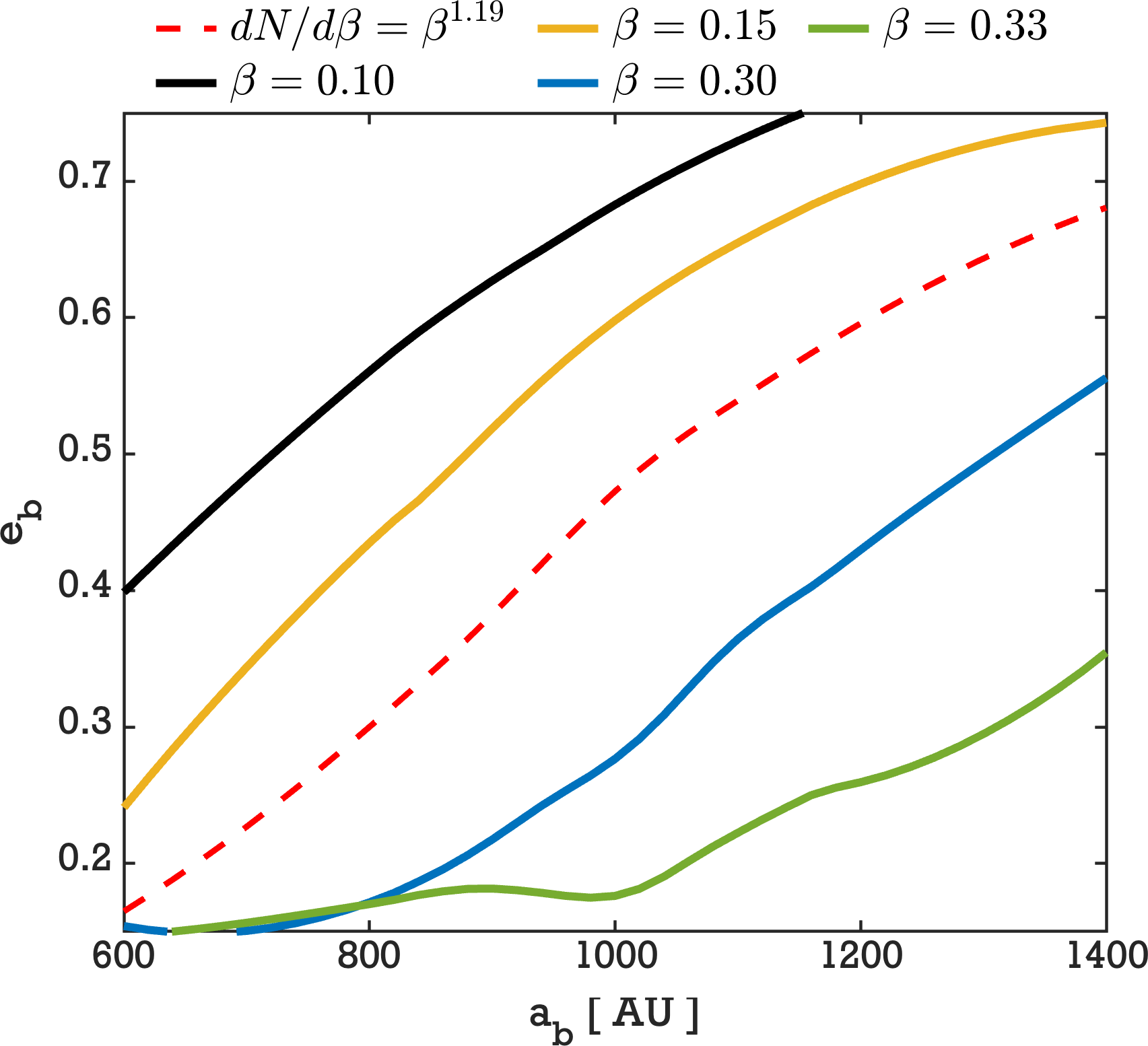}
\caption{Same as Figure \ref{disc_ecc_grav}, but allowing for radiation pressure. After spawning dust grains from planetesimal progenitors that evolved within our Laplace dynamical model, we compute the disc's eccentricity $\bar{e}_{\rm d}$ for a wide range of planetary semi-major axis and eccentricity covering the observational uncertainty intervals (Table \ref{table_system_parameters}). Different curves correspond to different values of $\beta$, the ratio between stellar radiation pressure and gravitational forcing. Prescribed $\beta$-values are randomly selected; while the distribution is sampled from the dust grain size distribution inferred in \citet{crotts2021deep}. Contours bound from below the area of the orbital parameter space in which the disc's eccentricity is pumped above 0.16.  
}
\label{disc_ecc_beta}
\end{figure}

Following the first approach, and allowing for variations of planetary semi-major axis and eccentricity within the uncertainty intervals inferred in \citet{Nguyen_2020} and listed in Table \ref{table_system_parameters}, we compute the global density-weighed disc eccentricity $\bar{e}_{\rm d}$ and contour its  surface in Figure \ref{disc_ecc_grav}. As expected, a closer and more eccentric planet generates a more eccentric disc. For almost half of the parameter space within the uncertainty intervals, the disc features minimal ($\bar{e}_{\rm d}\leq0.05$) to null eccentricity.  Meanwhile, a maximum disc eccentricity of $\bar{e}_{\rm d}=0.276$ can be attained with a planet having $(a_{\rm b}, e_{\rm b})= (600$ AU, $0.73)$. The nominal orbital parameters of the planet  from \citet{Nguyen_2020} are $(a_{\rm b}, e_{\rm b})= (850$ AU, $0.44)$, corresponding to $\bar{e}_{\rm d}=0.07$. Exciting the disc eccentricity to $\bar{e}_{\rm d}\approx0.1$ requires the planetary pericenter to be no farther than 415 AU. For $\bar{e}_{\rm d}\geq0.2$, the study constrains the planet to a narrow window of the parameter space with a maximum pericenter distance of 240 AU. It would be interesting to see whether future, better resolved, characterization of the disc will confirm these constraints or call for more physics into our modelling.

On a scale smaller than that of the gravitationally interacting planetesimals, dust particles are spawned and are propelled outwards by stellar radiation pressure \citep[][]{burns1979radiation,wyatt2008review}. Detailed modeling of this effect will take us too far afield. Instead, we post process our simulations, parametrizing the effect through the $\beta$-parameter, the ratio between radiation pressure and the gravitational forcing, which we take to be inversely proportional to the geometric cross section of dust grains. Thus on the disc scale, $\beta$ features a distribution that is related to the size distribution of dust grains, with smaller particles featuring more sensitivity to radiation pressure. As a function of parent planetesimal orbital elements (unprimed), the orbital elements of the dust particles (primed) are given by \citep{burns1979radiation}:
\begin{align}
     a^\prime &= \frac{a(1-e^2)(1-\beta)}{1-e^2-2\beta(1+e\cos f)},\\
     e^\prime &= \frac{\sqrt{e^2+2\beta e\cos f+\beta^2}}{1-\beta}, \\ 
     \omega^\prime &= \omega + \arctan\left(\frac{\beta\sin f}{e + \beta\cos f}\right), \\
     i^\prime &= i, \\
     \Omega^\prime &= \Omega,
\end{align}
where $f$ is the true anomaly of the parent planetesimal. Given the orbit averaged nature of our simulations, we retrieve the distribution of planetesimal true anomalies from a uniform distribution of mean anomalies via the Kepler equation. 

Allowing for radiation pressure on dust grains, we compute the eccentricity of the simulated disc that can be imaged in scattered starlight. We do so for different fixed values of $\beta$ corresponding to different dust grain sizes, which can in turn be observed at different imaging wavelengths \citep[e.g.,][]{hughes2018debris}. We also compute the disc eccentricity assuming the dust size distribution inferred in  \citet{crotts2021deep}. The authors obtained for the latter a power law with index $q= 3.19^{+0.11}_{-0.20}$, from which we infer $dN/d\beta \varpropto \beta^{1.19}.$ This distribution is only slightly different from that of a collisional cascade, which is characterized by $q\approx 3.5$ \citep{dohnanyi1969collisional}. Our $\beta$-distribution extends to a maximum value, $\beta_{\rm max}$, which marks a marginally bound orbit\footnote{To be sure, dust grains are likely to encounter HD 106906 b before hitting this limit, which would suggest a smaller upper-bound on $\beta$. This effect is worthy of consideration along with others ignored in this preliminary look at radiation pressure.}, and is defined for every parent planetesimal as:
\begin{equation}
\beta_{\rm max} = \frac{1-e^2}{2(1+e\cos f)} \approx 0.5 + \mathcal{O}(e).   
\end{equation}
In Figure \ref{disc_ecc_beta}, for each prescribed $\beta$ value or distribution, we mark the contour in $(a_{\rm b}, e_{\rm b})-$space that defines the parameter space required to generate a disc eccentricity $\bar{e}_{\rm d}\geq 0.16$. The latter value was obtained in \citet{crotts2021deep} by fitting the disc's spine to an eccentric ring. Evidently, the larger the value of $\beta$ the more propelled the dust grains are and the more eccentric the disc gets for a given pair of planetary parameters. For $\beta\geq0.33$, a planet with semi-major axis and eccentricity of any value within the uncertainty range can excite the disc's eccentricity above 0.16. For the $\beta-$distribution defined above ($dN/d\beta \varpropto \beta^{1.19}$), we are able to constrain $(a_{\rm b}, e_{\rm b})$ to within almost half of the parameter space (the area above the red contour curve in Figure \ref{disc_ecc_beta}). We find that the nominal values in  \citet{Nguyen_2020}, $(a_{\rm b}, e_{\rm b})= (850$ AU, $0.44)$, are within this constrained area. We finally note that for this nominal orbit, the resulting disc eccentricity is $\sim0.07$ when considering gravitational perturbations alone, and  $\sim0.19$ when accounting for dust particles with radiation pressure. Thus, the difference between the reported asymmetry using scattered light \citep{kalas2015direct,crotts2021deep}, and the axisymmetry in ALMA imaging \citep{kral2020survey} may very well be a natural outcome of radiation pressure.

\subsection{Inclinations again: Analogues and potential implications}

The simulated inclination profile of constituent planetesimals is a direct proxy for the vertical structure of the disc. Closer to home, the Kuiper belt has been the focus of numerous studies attempting to model its vertical structure \citep{brown2001inclination,elliot2005deep,kavelaars2009canada}. Such a characterization is crucial to better constrain scenarios for the belt's formation and issuing dynamical evolution  \citep{gomes2003origin,levison2003formation,nesvorny2015jumping}. Through such studies, it has been established that the Kuiper belt features two distinct populations, one with low inclination objects (the cold population), and another with high inclination objects (the hot population). Those populations are often modelled by the sum of two overlapping Gaussians \citep{brown2001inclination}: a narrow Gaussian for the cold population, and a broader Gaussian for the hot one, with possible refinements on the former \citep[e.g.,][]{petit2011canada}. A similar decomposition has been attempted for discs beyond our solar system, particularly in the $\beta$ Pictoris system where \cite{matra2019kuiper} again identified hot and cold planetesimal populations. 

\begin{figure}
\centering
\includegraphics[width=\linewidth]{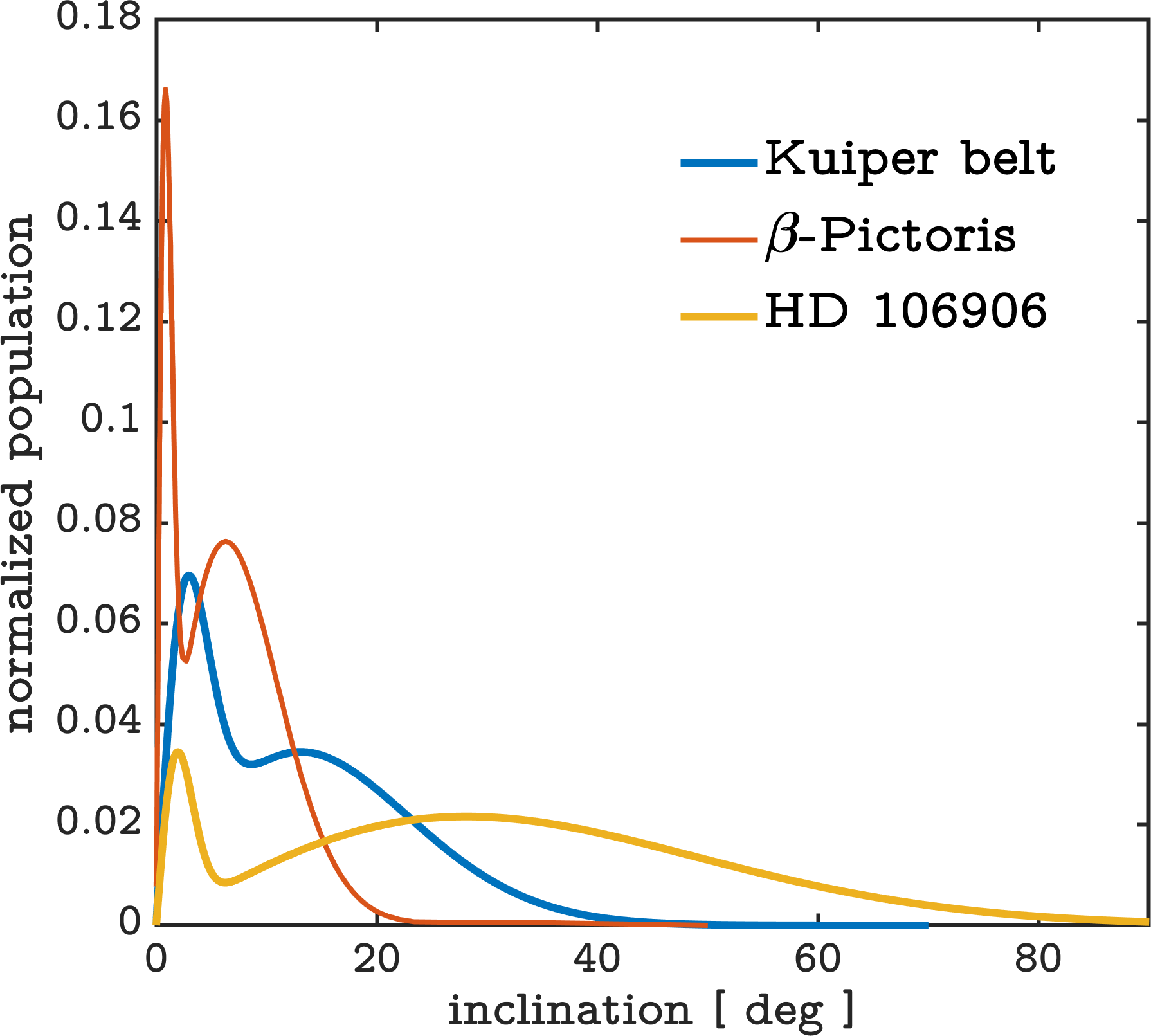}
\caption{Inclination distributions as derived and modelled from observations of the Kuiper belt \citep{brown2001inclination}, ALMA observations of $\beta$ Pictoris \citep{matra2019kuiper}, and our dynamical simulations of HD 106906, captured at the system's age (13 Myr). Each of the  three distributions is fitted to two overlapping Rayleigh distributions describing a high inclination hot planetesimal population, and a low inclination cold population.}
\label{discs_distributions}
\end{figure}

We plot in Figure \ref{discs_distributions} the population density for the inclination distribution of the Kuiper belt and $\beta$ Pictoris.  We further include the inclination distribution of our simulated HD 106906 system at its current age. We use the same surface density distribution used before and described in Appendix \ref{Appendix_Maps}  (i.e., \eq{eq:Sigmad_PL} with $p=0.5$). Our simulated HD 106906 disc features a similar distribution, with distinct cold and hot populations. This bimodal distribution is an inherent aspect of Laplace dynamics as elucidated thus far. The cold population is predominantly attributed to the inner part of the disc living in the binary-dominated regime; whereas the hot population mainly is mainly the planet-dominated regime. We fitted this distribution to two overlapping Rayleigh distributions with a mean and standard deviation of $1.89$ and $1.24$ degrees for the cold population, and $28.08$ and $18.39$ degrees for the broader hot population

Accounting for uncertainties on the planetary orbit as we did in Section \ref{variations_planet_rL}, we noted significant shifts in the Laplace radius which can alter this bimodal distribution significantly. An extreme of course is a system where the perturbations due to the planet are negligible, and this occurs when the planet's actual semi-major axis and eccentricity are towards the upper and lower ends of their estimated range, respectively (see Figure \ref{fig:radius} and last panel of Figure \ref{surface_density_varying_rL}). In this case, the dominance of the inner binary over the planet leads to the sculpting of a single, rather than two, Gaussian distribution in inclination centered around $0^{\circ}$. In the opposite case where the planet overshadows the binary's gravitational field, the vertical distribution is expected in the longterm to show a single peak which is now centered around large values, as seen when we mute the inner binary in Figure \ref{Histograms}. Thus, both the number and positions of peaks in the vertical distribution of planetesimals could help constrain the planet's orbit. Their mapping will have to await future  high-resolution ALMA observations, and can then indirectly complement future RV and astrometric measurements of HD 106906 b.  (see also Section \ref{variations_planet_rL}, where we discuss $\bar{e}_{\rm d}$)\footnote{The axisymmetric structure of the disc reported in \cite{kral2020survey} and \cite{meredith_106906} favors a Laplace dynamical regime in which the stellar binary dominates the full disc, which would in turn favor a dominant cold population in any future sufficiently resolved characterization of the disc's vertical structure.}.

Finally, and conversely, future refinements in the mass and orbital parameters of HD 106906 b could provide further support [or lack thereof depending on the findings] for the dominant role of the binary in this specific system. This represents a promising avenue, especially in light of the advent of JWST which could directly image planets of $\gtrsim 0.1 M_{\rm J}$ at separations of $\sim 100$ au and beyond \citep[e.g.,][]{carter2021direct}.

Before moving on, we note that our predictions are, to some extent, model-dependent via the prescribed disc's innermost and outermost radii  (Appendix \ref{Appendix_Maps}), together with the mass distribution  controlled by $p$ (Eq. \ref{eq:Sigmad_PL}). We have, however, confirmed that this dependence is weak  for all astrophysically-motivated values of $p$ (i.e., $ 0 < p < 2$) as long as $a_{\rm in} \ll a_{\rm out}$. Finally, we emphasise that our results derive from a collisionless model, a limitation that must be assessed before making any definitive conclusions. Collisional activity within the disc may have a significant -- and potentially dominant -- influence on the inclination distribution of the planetesimals. While this is beyond the scope of our current work, we expect collisions to predominantly deplete the innermost planetesimals, i.e., those with $a \lesssim r_{\rm L}$, and thus affect the peak of the cold population in Figure \ref{discs_distributions}. This is so because relative velocities of planetesimals would be higher in the inner parts than in the outer parts (recall that $v_{\rm rel} \propto v_{\rm K} \propto r^{-1/2}$, where $v_{\rm K}$ is the Keplerian speed). The extent of depletion further depends on the micro-physics governing the collisional activity \citep{wyatt2008review}, e.g., the strength of the planetesimals, the distribution of sizes, as well as the maximum size. There is clearly a need for a more sophisticated analysis once and when the vertical distribution of the HD 106906 disc is sufficiently well-resolved by ALMA. 

\subsection{A variety of vantage points}\label{section_vantage_points}

\begin{figure*}
\centering
\includegraphics[width=\linewidth]{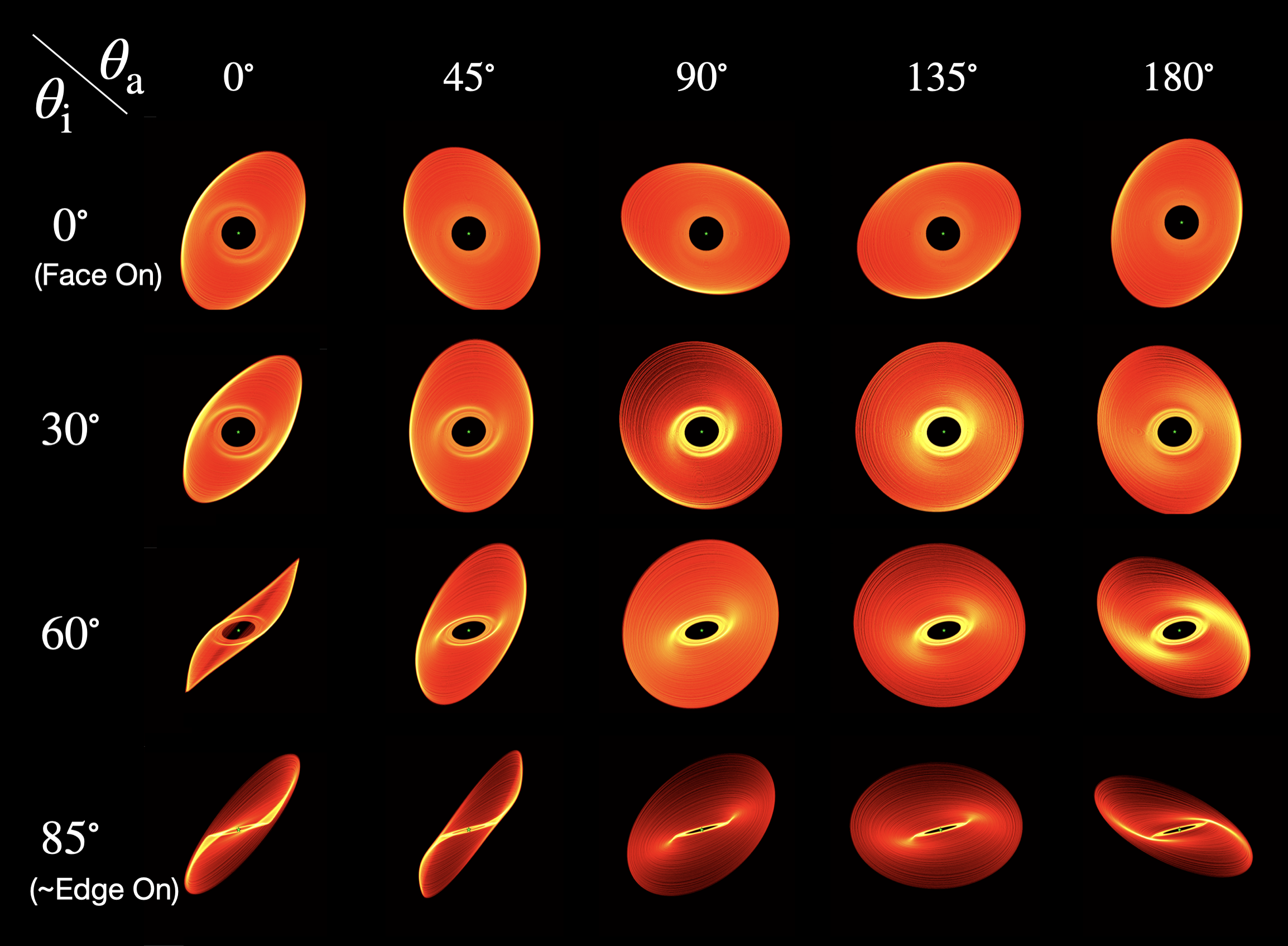}
\caption{Synthetic maps of surface density corresponding to HD 106906-like discs at $t = 13 $ Myr as viewed from a variety of vantage points. Results are shown as a function of line-of-sight inclination angle $\theta_i$ to the observer and azimuthal angle $\theta_a$; see Section \ref{section_vantage_points} for further details. Each image is constructed following the procedure described in Appendix A, has 800x800 pixels, and is scaled individually to bring out faint details.
The green star symbol in each panel represent the location of the inner binary's centre of mass. One can see the presence of two-armed spiral pattern for discs which are nearly face-on, i.e., with $\theta_i \lesssim 90^{\circ}$, and a warp for nearly edge-on discs, i.e., for $\theta_i \sim 90^{\circ}$.}
\label{Gallery_binary_on}
\end{figure*}

Motivated by the fact that HD 106906 is seen nearly-edge on, we have thus far focused on describing the vertical structure imprinted by the Laplace surface dynamics (e.g., Figure \ref{surface_density_maps_binary_on}). For completeness, we now probe the appearance of the HD 106906 disc at its current age when viewed from a variety of vantage points. We do so to exemplify observables of Laplace surface dynamics for similar such dynamical configurations, over a range of potential orientations. This is illustrated in  Figure \ref{Gallery_binary_on}, where we plot the surface density of our simulated HD 106906 disc as viewed on the sky plane for various combinations of $\theta_{\rm i}$ and $\theta_{\rm a}$ (see Appendix \ref{Appendix_Maps}). Note that each of the panels in this figure is normalised individually to bring out the fainter features, and $\theta_{\rm l}$ is kept fixed\footnote{Recall that $\theta_{\rm l}$ measures the angle in the sky-plane from the positive $X$-axis to the disc's ascending node on the sky (see Appendix \ref{Appendix_Maps}), and thus varying its value will not affect the disc structure.} at the fiducial value of $13.7^\circ$.

Looking at Figure \ref{Gallery_binary_on}, one can see that one of the common features evident in each of the displayed panels is the little or no evidence of any stellocentric offset.  In other words, the stellar binary's centre of mass, denoted by a green star, coincides with the apparent geometric center of the disc's inner cavity, as can be seen most prominently for the nearly face-on discs, i.e., $\theta_{\rm i} < 30^{\circ}$. This simply is a restatement of the fact that the stellar binary's perturbations tame the eccentricity excitation by the external planet, otherwise expected to be significant in the inner parts of the disc (Figure \ref{evolution_integration}, see also \citealp{nesvold2017hd}). 

Another common feature in many of the views displayed in Figure \ref{Gallery_binary_on} is the presence of a two-armed spiral pattern.  This pattern, as can be seen more readily for discs viewed with $\theta_i \lesssim 90^{\circ}$, starts off within the disc, i.e., $a_{\rm in} \ll r \ll a_{\rm out}$, and extends out to the outer disc parts. This can be understood by noting that by this time, i.e., $t = 13$ Myr, the stellar binary has homogenised the apsidal angles of planetesimals, together with their longitudes of ascending nodes in  the inner disc parts i.e., $a \lesssim r_{\rm L}$, while those at $r \gtrsim r_{\rm L}$ have not had completed one complete precession cycle   (see Figures \ref{fig_timescales} and \ref{evolution_integration}). Thus, by and large, the spiral pattern demarcates the regions of dominance by the inner binary and the exterior planet.  Note that while this spiral pattern is expected to be transient in nature, and thus should eventually wrap onto itself given enough time, Figure \ref{evolution_integration} indicates that the pattern may survive for relatively long times, i.e., $t \sim 100$ Myr\footnote{This can be also understood by consulting Figure \ref{fig_timescales} which shows that the precession timescale for $a \gtrsim r_{\rm L}$ has a weak dependence on semi-major axis.}. Note that similar spiral patterns were observed by \citet{Beust-2017} who reported on HD 106906 debris disc dynamics, allowing for the binary and a co-planar planet in a study which is mainly concerned with a dynamical origin for that planet.

\subsection{Generality of Laplace surface dynamics}

By applying our model to HD 106906, we have shown that Laplace surface dynamics can play a significant role in shaping this system's debris disc for a wide range of system parameters -- see e.g. Figure \ref{surface_density_varying_rL} and Section \ref{section_dynamics_around_LS}.  In the competition between an equal-mass stellar binary interior to the disc and a super-Jupiter planetary companion exterior to it, and over a wide range of triple parameters, the disc was shown to sustain a winding spiral within a twisting warp both having distinctive kinematic and photometric signatures as discussed above. Within our multiply averaged framework for hierarchical systems, we can handle a whole slew of configurations, from unequal mass binaries of varying separation, including a Sun-Jupiter pair, to wide companions ranging from the planetary to the stellar. 

Essentially, it all comes down to the location of the Laplace radius as captured in Eq.\ref{laplace_radius}. With a solar mass external companion, the early spirals and warps of Laplace surface dynamics are pushed to inner, typical planet forming regions, with debris in the hundred AU range now fully exposed to the tug of this stellar companion. On the other hand, Sun and outer solar system can work with a Neptune-mass ninth planet to sculpt a Laplace surface in trans-Neptunian realm, i.e. in the few hundred AU range; this range shifts inward (outward) with decreasing (increasing) companion distance, and/or larger (smaller) companion eccentricity. Suffice it to say that a sweep over system parameters will readily reveal the process described here to be generic to debris discs over a wide range of semi-major axes in triples of various configurations. 

Of interest in this context is the case of HD 100453 which features an inner binary, a known stellar companion, and a debris disc with a warp, then a gap and a few Jupiter mass planet hypothesized in that gap, and a substantial gaseous component \citep{dong2015m,van2019alma}. Observations reveal spirals and a warp with a break around 27 AU \citep[e.g.,][]{wagner2015discovery}. Associating the break with a Laplace radius which is controlled by inner planet, and outer stellar companion, we find that a 5 Jupiter mass planet within the gap\footnote{ This is as good a spot as any to raise the possibility of a hitherto undetected few Jupiter-mass planet within HD 106906's inner cavity. Such a planet would sculpt the disc's inner edge, as it couples to the binary in dictating debris dynamics by shifting $r_{\rm L}$ further out, among other potential effects. Further consideration of this possibility is best left till when future refined observations warrant it.} will enforce a Laplace transition around 27 AU, where observations suggest it to be.This straightforward estimate is consistent with the conclusion of the extensive modeling and simulations in \cite{nealon2020spirals}, all the while necessitating further elaboration to allow for the gaseous component. 

This state of affairs raises concerns about the dynamical modeling of debris discs in general which often considers either an inner or an external perturber \citep[e.g.,][]{wyatt1999observations,chiang2009fomalhaut, pearce2014dynamical,lee2016primer,  nesvold2017hd}, and proceeds to fit for its mass, location and/or eccentricity and inclination. Such systems, and associated modeling, are worth revisiting for sure, allowing for the combined effect of inner and outer companions (observed or presumed). 

We note in passing that for the equal mass binary in HD 106906, octupolar perturbations cancel out, leaving the included quadrupolar contribution as the dominant term (Section \ref{section_dynamics_model}). However, a binary composed of unequal-mass stars, if and when on an eccentric orbit, can further disturb test-particle dynamics at the octupolar level potentially carving the axisymmetric sub-region under its influence. Also, and depending on the strength of the octupole, and the spatial extent of the debris disc, it could also introduce another level of competition between the inner and outer components of the planetary system further shaping the Laplace surface warp. It is straightforward to generalize the analysis above to such configurations, an extension which we defer to the notorious category of future works. 

We further note that debris disc mass and resulting self-gravity, which we ignored in this already challenging first look at the problem, could alter the dynamics reported herein. Disc self-gravity was already highlighted as a key component of generalized co-planar Laplace dynamics by \cite{Sefilian2019} who showed that the combined effect of disc self-gravity with an inner quadrupole can freeze orbital precession over a range of semi-major axes in a disc of few Earth masses and moderate eccentricity. It is not difficult to estimate \citep{Sefilian2019, Sefilian21} that an axisymmetric disc of mass $M_{\rm d} = 20 M_{\earth}$\footnote{For \cite{meredith_106906}, the observed 1.3 mm flux indicates a $\sim 10M_{\earth}$ reservoir of planetesimals, an estimate which surely depends on the assumed size distribution and the maximum size of planetesimals \citep{krivov_wyatt}.} yields an apse precession rate of $\approx - 0.09$ ${\rm Myr}^{-1}$ (hence a period of  $\approx 70$ ${\rm Myr}$) around $70$ AU, a rate which, as shown in Figure \ref{fig_timescales}, is comparable in magnitude to that induced by the combined effect of binary and planet at that location. The challenge here is to allow for disc self-gravity over a warping, precessing geometry which is induced by an external eccentric and inclined companion. We defer full consideration of this potentially important effect to future work, and simply note for now that a massive disc, which contributes negatively to apse precession, would usher an inward shift in the Laplace radius, all else remaining fixed. Such a shift would leave the disc exposed to twists and warps now unfolding in the combined field of binary, planet, and  disc! By insisting on the disc being maintained in an axisymmetric state, we would then need to revisit Sec.\ref{variations_planet_rL} constraints on the planet's orbit to make sure that the effective Laplace radius is once again beyond the disc's outer edge: a game worth playing for sure\footnote{The same is true for conclusions reached by \cite{meredith_106906} assuming a disc of test particles.}!

We close this section by emphasising that some of the above variations have already been explored around Solar system planetary satellites \citep{ward1981orbital,saillenfest2021past}, and further considered following the work of \cite{Tremaine} for exo-planetary systems with binary companions, allowing for planet migration, and disc self-gravity as and when called for \citep[e.g.,][]{zanazzi2016extended,spalding2020stellar,speedie2020structure}. Here, we are simply emphasising the same for debris discs in configurations that allow for substantial companion eccentricity and inclination.

\section{Summary and outlook}

Much was recognized in common between HD 106906 and our solar system. Both have a prominent debris disc, with evident inner quadrupolar forcing (HD 106906's stellar binary, the solar system's planets). HD 106906 features a directly imaged, and astrometrically constrained distant planet, surmised to disturb a young planetesimal/debris disc, which is guessed at through the dust cloud that shrouds it but otherwise inaccessible to our current observatories. Our solar system features an exquisitely mapped debris disc with Trans-Neptunian Objects showing signs of clustering dynamically attributed to hypothesized and yet to be imaged distant planet. 

We had recently generalized the Laplace surface dynamics toolbox to apprehend and successfully recover key ingredients in support of Planet 9 shepherding of TNOs \citep{farhat2021laplace}. With the analogies outlined here, it was not surprising for this toolbox to be particularly handy (with minor modifications) to address the relatively transient dynamics of the debris disc in HD 106906, which is perturbed by the inner binary and a directly imaged distant super-Jupiter. We had developed this machinery to map out generalized Laplace equilibria and neighboring dynamics with a view to identifying islands that can harbor a fairly evolved debris population in the solar system. Here, we deployed it to map out transient dynamics shaping current observations of HD 106906 debris disc. We further extended analysis, and simulations to allow for a range of parameters, and system ages, permitting conclusions for a broader ensemble of exo-planetary systems.

We revisited  HD 106906 having convinced ourselves with a back-of-the-envelope calculation that the inner binary is a significant player in the disc’s dynamics as the outer planetary companion. For the nominal orbital parameters of the system, the comparable effects of the inner binary and the outer planet impose an inevitable sharp break of dynamical regimes, leaving behind an inner flat and circular region that warps into a vertically thicker and asymmetric belt in the outskirts, with spiral arms anchored to the edges of the warp. This well defined structure gives rise to a bimodal vertical distribution that mimics those observed in the Kuiper belt and $\beta$ Pictoris. In light of contrasting observational inferences on the disc's asymmetry \citep[e.g.,][]{kral2020survey,crotts2021deep}, we further studied the disc over the full range of system parameters, within uncertainties, and confirmed that the disc can range from being fully dominated by the inner binary, thus forced to live axi-symmetrically in the binary's plane, to being significantly so, with the Laplace radius providing the demarcation line. A lingering uncertainty involves the planet's orbit, so we mapped the general dependence of the disc's asymmetry on the planet's parameters with and without radiation pressure. Our analysis shows that, for the nominal orbital parameters of the planet \citep[][]{Nguyen_2020}, the disc would appear eccentric in scattered light and nearly circular in ALMA, thus explaining the discrepancy. It will be interesting to confront our conclusions with improved constraints on the orbit of HD 106906 b, along with more resolved images of the disc.  We then closed the loop on our dynamical study by probing the phase space structure of the disc. We learned that initially cold discs are destined to evolve on a robust phase space structure, specifically a separatrix which carries planetesimals on excursions of inclination and eccentricity which grow in amplitude with increasing semi-major axis. We  recovered a closed form solution for this separatrix, and noted how this structure and neighboring dynamics help capture the long-term evolution of similar such discs.

The hypothesized Planet 9 motivated the development of a toolbox which was deployed on HD 106906 debris disc. From this work's perspective on debris disc dynamics, we are now tempted to revisit the Kuiper belt in the context of a dynamically evolving solar system and in the presence of a primordial ninth Planet. We envisage it scultped with double spirals and warps, all the while the Laplace radius migrates with the outer planets: what role, if any, does this migration play in the clustering of TNOs around stable high inclination and eccentricity islands? Can the observed double peak in the inclination distribution be partly attributed to a warping Laplace surface in the presence of a distant outer perturber?

In this work, we assumed that HD 106906 b was around in its current position at the get go. This is certainly the simplest choice when concerned with secular dynamics of the debris disc. The outcome of our calculations can then be confronted with detailed observations to decide whether disc structure is consistent with the gentle secular forcing by a primordial distant planet formed in-situ \citep{maury2019characterizing,jennings2021primordial}, or whether it holds signs of more violent disruption that could have resulted from a planet formed close in then disrupted in the course of migration then resonance with the inner binary \citep{Beust-2017}. 

Of course, there is the further complication of the primordial gas on one hand and the collisional evolution of the debris on the other. The interplay between secular debris dynamics and the gaseous component of protoplanetary discs around binaries was clarified by \cite{rafikov2014planet}, then further generalized to nonlinear regimes by \cite{Sefilian2022} [drawing on \cite{Sefilian_Thesis}]. In the process we learned: that gas gravity can dramatically alter collision rates; that secular resonances can sweep through the disc in the course of gas dissipation,  altering debris disc eccentricity (and surely inclination) distributions; and that such sweeping together with gas drag can significantly sculpt debris distributions.

One naturally wonders how all of this carries over to HD 106906 Laplace surface dynamics with a disc that is fairly young, apparently massive enough to be self-gravitating, and has surely emerged from a primordial now largely dissipated massive gaseous component. The question resonates with work on warping self-gravitating discs by \cite{toomre1983theories}, as well as the early and insightful meditation of \cite{ward1981orbital} on the role of formative gas in the inclination of Iapetus. Both studies, with separate motivations, emphasise the role of disc gravity in shaping the structure of the warp in the presence of external forcing. 

When it comes to HD 106906, we may very well be justified in assuming an initial debris disc which is cold and co-planar with the binary. This is so because, as it happens, a million year gas dissipation timescale is at worst comparable, at best faster than secular timescales around the system's Laplace radius. Still, and for reasons discussed above, HD 106906 invites various critical generalizations of Laplace surface dynamics to allow for a dissipating gaseous component, disc self-gravity and collisional evolution. Such generalizations will help us recover physically secure initial conditions for our calculations, then draw the implications of disc self-gravity and collisional evolution for the currently observed structure. More generally, and perhaps more importantly, they will give us the pleasure of weaving yet another thread in the rich and layered tapestry initiated by Laplace more than two centuries ago. 


\section*{Acknowledgements}
The work resulted from a back-of-the-envelope calculation which came together around a refined meal at Badguer, Bourj Hammoud, Lebanon. J.R.T. wishes to express his gratitude to its hostess, Madame Arpi Mangassarian, for sustaining an island of warmth and serenity in the midst of utter collapse. The authors are grateful to the referee, Herv\'e Beust, for insightful comments which helped improve the manuscript considerably. 
A.A.S. is supported by the Alexander von Humboldt Foundation through a
Humboldt Research Fellowship for postdoctoral researchers.
A.A.S. also acknowledges partial financial support from ESO/Gobierno de Chile during the early stages of this work.

\section*{Data Availability}

The data underlying this article will be shared on reasonable request
to the corresponding author.



\bibliographystyle{mnras}




\appendix

\section{The Secular Hamiltonian}\label{Appendix_secular}
We are interested in the motion of a massless planetesimal which evolves in the combined gravitational field of a central stellar binary and an outer massive planetary companion. We provide here details on the derivation of the orbit-averaged Hamiltonian of Eq.\ref{basic_Ham} which drives the dynamical evolution of this planetesimal via Eqs.[\ref{eom_j}, \ref{eom_e}]. Following notation in the body of the text, the central binary components have masses $m_{\rm A}$ and $m_{\rm B}$ and total mass $m_{\rm AB}=m_{\rm A}+m_{\rm B}$. We denote their instantaneous barycentric position vectors by $\boldsymbol{r}_{\rm A}$ and $\boldsymbol{r}_{\rm B}$ with $\boldsymbol{r}_{\rm AB}= \boldsymbol{r}_{\rm A}-\boldsymbol{r}_{\rm B}$. We further denote by $\boldsymbol{r}$ the instantaneous position vector of the planetesimal relative to the barycenter of the binary. Similarly, we prescribe  for the outer planet the mass $m_{\rm b}$ and the position vector $\boldsymbol{r}_{\rm b}$. With these definitions, debris dynamics is governed by the Hamiltonian 
\begin{equation}
    H = \frac{1}{2}\dot{\boldsymbol{r}}^2 -G \left(\frac{m_{\rm A}}{|\boldsymbol{r}-\boldsymbol{r}_{\rm A}|} + \frac{m_{\rm B}}{|\boldsymbol{r}-\boldsymbol{r}_{\rm B}|} + \frac{m_{\rm b}}{|\boldsymbol{r}-\boldsymbol{r}_{\rm b}|}\right).
\end{equation}
Denoting by $\mu$ the mass ratio $m_{\rm B}/m_{\rm AB}$, the position vectors become $\boldsymbol{r}_{\rm A}=\mu\boldsymbol{r}_{\rm AB}$ and $\boldsymbol{r}_{\rm B}= (\mu-1)\boldsymbol{r}_{\rm AB}$, allowing us to rewrite the Hamiltonian as
\begin{equation}
    H = \frac{1}{2}\dot{\boldsymbol{r}}^2 -G \left(\frac{m_{\rm A}}{|\boldsymbol{r}-\mu\boldsymbol{r}_{\rm AB}|} + \frac{m_{\rm B}}{|\boldsymbol{r}-(\mu-1)\boldsymbol{r}_{\rm AB}|} + \frac{m_{\rm b}}{|\boldsymbol{r}-\boldsymbol{r}_{\rm b}|}\right). 
\end{equation}
One then expands the potential in terms of distance ratios, $r_{\rm AB}/r$ and $r/r_{\rm b}$ respectively, and the usual Legendre polynomials, $P_{\ell}$. For instance: 
\begin{equation}
    \frac{1}{|\boldsymbol{r}-\boldsymbol{r}_{\rm b}|} = \frac{1}{r_{\rm b}}\sum_{\ell=0}^{\infty}\left(\frac{r}{r_{\rm b}}\right) ^\ell P_{\ell}(\cos\gamma),
\end{equation}
where $\cos\gamma = (\boldsymbol{r}\cdot\boldsymbol{r}_{\rm b})/(rr_{\rm b})$. Such expansions are reliable provided the system's architecture is sufficiently hierarchical for the series to converge. Given the nominal orbital architecture of the studied system, HD\,106906, described in Table \ref{table_system_parameters}, we truncate the expansions at the octupolar order, i.e. at order 3 in $r_{\rm AB}/r$ and $r/r_{\rm b}$ respectively. The Hamiltonian then reads
\begin{equation}\label{H_insta}
    H = H_{\rm K} + \Phi_{\rm AB,\,quad}+ \Phi_{\rm AB,\,oct}+ \Phi_{\rm b,\,quad}+ \Phi_{\rm b,\,oct},
\end{equation}
where we have defined
\begin{align}
 & H_{\rm K} = \frac{1}{2}\dot{\boldsymbol{r}}^2 - G\frac{m_{\rm AB}}{r},\label{ham_osc_1}\\
& \Phi_{\rm AB,\,quad}= -\frac{G}{2r^3}\frac{m_{\rm A}m_{\rm B}}{m_{\rm AB}}\left[\frac{3(\boldsymbol{r}\cdot\boldsymbol{r}_{\rm AB})^2}{r^2}-r_{\rm AB}^2\right],\\
& \Phi_{\rm AB,\,oct}= -\frac{G}{2r^5}\frac{m_{\rm A}m_{\rm B}}{m_{\rm AB}^2}(m_{\rm A}-m_{\rm B})\left[ \frac{5(\boldsymbol{r}\cdot\boldsymbol{r}_{\rm AB})^3}{r^2}-3(\boldsymbol{r}\cdot\boldsymbol{r}_{\rm AB})r_{\rm AB}^2\right ],\label{binary_oct}\\
&\Phi_{\rm b,\,quad} = -\frac{G m_{\rm b}}{2r_{\rm b}^3}\left[\frac{3(\boldsymbol{r}\cdot\boldsymbol{r}_{\rm b})^2}{r_{\rm b}^2}-r^2\right],\\
& \Phi_{\rm b,\,oct}= -\frac{Gm_{\rm b}}{2r_{\rm b}^5}\left[ \frac{5(\boldsymbol{r}\cdot\boldsymbol{r}_{\rm b})^3}{r_{\rm b}^2}-3(\boldsymbol{r}\cdot\boldsymbol{r}_{\rm b})r^2\right ].\label{ham_osc_fin}
\end{align}
It is evident from \eq{binary_oct} that the octupolar contribution of the inner binary vanishes when the binary components have equal masses, which is almost the case of HD 106906. We therefore ignore this contribution in our study. To proceed further, we rely on the reference frames introduced in the main text and defined by the triads $(\boldsymbol{\hat{n}},\boldsymbol{\hat{u}},\boldsymbol{\hat{v}})$, $(\boldsymbol{\hat{n}}_{\rm AB},\boldsymbol{\hat{u}}_{\rm AB},\boldsymbol{\hat{v}}_{\rm AB})$, and $(\boldsymbol{\hat{n}}_{b},\boldsymbol{\hat{u}}_{\rm b},\boldsymbol{\hat{v}}_{\rm b})$. Bound planetesimal motion is then given by 
\begin{equation}
    r=\frac{a(1-e^2)}{1+e\cos(f)}\,\,, \hspace{1cm} \vec r=r (\cos(f) \,{ \hat{u}} +\sin(f) \,{ \hat{v}})
\end{equation}
where $a$ and $e$ are the planetesimal's semi-major axis and eccentricity, and $\nu$ is its true anomaly relative to $\boldsymbol{\hat{u}}$. 

Secular motion is then captured by  averaging Hamiltonian contributions in Eqs. (\ref{ham_osc_1}-\ref{ham_osc_fin}), first over $P$, the orbital period of the planetesimal on its osculating Keplerian orbit, then over the  orbital period of the outer planet. In doing so, it is useful to keep in mind the following differential relations
\begin{equation}
\frac{dt}{P}=\frac{rdE}{2\pi a}=\frac{r^2d f}{2\pi a b}=\frac{dM}{2\pi},
\end{equation}
where $b = a \sqrt{1-e^2}$ is the semi-minor axis, \textit{E} the eccentric anomaly, and \textit{M} the mean anomaly, with analogous expressions for the outer planet. In performing those averages, we rely on the evident separation of timescales between the orbital periods of the inner stellar binary, the planetesimal, and the planetary companion, while ignoring possible, higher order, mean motion resonances. We then recover expressions for the orbit-averaged multipoles, parameterized by the normalized angular momentum vector, $\boldsymbol{j} =\sqrt{1-e^2} {\boldsymbol{\hat{n}}}$ and the Lenz vector $\boldsymbol{e}  = e {\boldsymbol{\hat{u}}}$, via a straightforward though somewhat laborious exercise which is now well documented in various publications on hierarchical triples \citep[e.g.,][]{Tremaine, correia2011tidal, hamers2020secular,farhat2021laplace}.
\section{Constructing Maps of Surface Density}
\label{Appendix_Maps}

Here, we describe our recipe to construct maps of disc surface density using the distribution of the planetesimal orbital elements. The procedure we follow is generally similar in spirit to that outlined in \citet{Sefilian21} for razor-thin discs (see Appendix C therein), but now generalised to account for orbital inclinations and viewing angles. 

Maps are built with $N=2410$ planetesimals drawn from a uniform distribution in the semi-major axis $a$ from the $24100$ planetesimals constituting our simulated debris disc. We assign a mass $m_i$, $i = 1, .., N$, to each planetesimal in the sample in such a way that the initial disc surface density $\Sigma_{\rm d}^{t=0}$ varies with radius $r$ -- or equivalently with  $a$, since planetesimals are initiated on circular and coplanar orbits (Section \ref{section_dynamics_around_LS}) -- following a power-law profile given by:
\begin{equation}
    \Sigma_{\rm d}^{t=0}(a) = \Sigma_0 \left( \frac{a_{\rm out}}{a} \right)^p , 
    \label{eq:Sigmad_PL}
\end{equation}
for $a_{\rm in} \leq a \leq a_{\rm out}$. Unless otherwise stated, we use $p=0.5$, and we fix the innermost and outermost particles to have $a_{\rm in} = 30 {\rm au}$ and $a_{\rm out} = 150 {\rm au}$. Given that the semi-major axis distribution remains constant within the secular regime, we make use of the relationship \citep{statler2001}:
\begin{equation}
 dm(a) = 2 \pi a  \Sigma_{\rm d}^{t=0}(a) da,
\end{equation}
and we compute the amount of mass per unit semi-major axis for a given initial surface density (Equation \ref{eq:Sigmad_PL}).

After evolving the planetesimal distribution in time, at a given snapshot of our simulations, we smear the mass of each planetesimal over its orbit, i.e., into an eccentric ring. This is done by spawning $N_{\rm p} = 5 \times 10^{4}$ new particles for each of the $N$ parent planetesimals, such that the new particles share the same orbital elements as the parent planetesimals, have masses $m_i/N_{\rm p}$, and  mean anomalies $M$ that are randomly selected from the $[0, 2\pi]$ range. This procedure guarantees that each ring is characterised with a non-uniform linear density reflecting the fact that particles on eccentric orbits spend more time at their apocentre than at their pericentre. Next, we compute the eccentric anomaly $E$ for each of the considered particles, which in total sum up to $N_{\rm t} = N \times N_{\rm p}$, using Kepler's equation: 
\begin{equation}\label{kepler_equation}
    M = E - e \sin E .
\end{equation}
This allows us to obtain the position of each particle via standard formulae:
\begin{eqnarray}
x &=& r \left( \cos \Omega \cos(\omega+f) - \sin \Omega \sin(\omega+f) \cos i \right) ,  
\nonumber \\
y &=& r \left(\sin \Omega \cos(\omega+f) + \cos \Omega \sin(\omega+f) \cos i  \right) ,  
\nonumber \\
z &=& r \sin(\omega + f) \sin i ,
\label{eq:XYZ-planetesimals}
\end{eqnarray}
where, as usual, one has $r= a (1-e\cos E)$. For ease of description in what follows, we refer to the coordinate system $(x,y,z)$ as the ``disc-plane'' system.

Next, we transform from the ``disc-plane'' system to the ``sky-plane'' coordinate system, denoted by $(X,Y,Z)$. We do this in such a way that the sky plane is represented by the $(X,Y)-$plane, with the positive $X-$ and $Y-$axes pointing west and north, respectively, and the positive $Z-$axis pointing along the observer's line of sight. This transformation can be done via:
\begin{equation}
\begin{pmatrix} 
X  \\  Y \\ Z
\end{pmatrix}
= \mathcal{R}_z(\theta_{\rm l}) \cdot \mathcal{R}_x(\theta_{\rm i}) \cdot \mathcal{R}_z(\theta_{\rm a})
\begin{pmatrix} 
x \\ 
y \\
z
\end{pmatrix}
,
\label{eq:XYZ-sky}
\end{equation}
where $\mathcal{R}_x(\theta)$ and $\mathcal{R}_z(\theta)$ represent counter-clockwise rotations about the $x-$ and $z-$axes such that:
\begin{equation}
 \mathcal{R}_x(\theta) =
    \begin{pmatrix}
   1 & 0 & 0 \\
   0 & \cos \theta & -\sin \theta \\
   0 & \sin \theta & \cos \theta 
    \end{pmatrix}
    , ~~~ 
     \mathcal{R}_z(\theta) =
    \begin{pmatrix}
   \cos \theta  &  -\sin \theta  & 0 \\
   \sin \theta & \cos \theta & 0 \\
   0 & 0 & 1
    \end{pmatrix}
    .
\end{equation}
In Equation (\ref{eq:XYZ-sky}), the angle $\theta_{\rm l}$  is the angle in the sky plane, i.e., in $(X,Y)$, measured from the positive $X-$axis to the disc's apparent ascending node on the sky. It can be determined by the apparent major axis of the disc, which, for HD 106906, has a position angle ${\rm P.A.} = 283.7^{\circ}$ \citep{kalas2015direct}. Given that the pericenter side of the HD 106906 disc appears to be towards the East-West direction, one then has $\theta_{\rm l} = {\rm P.A.} + \pi/2$ (i.e., anti-clockwise from North)\footnote{A similar result could be obtained using the value of $PA = -74.7^{\circ}$ of \citet{Olofsson22}, in which case $\theta_{\rm l} = -74.7 + 90 = 15.3^{\circ}$.}. The angle $\theta_{\rm i}$, on the other hand, is a measure of the angle between the normals to the sky- and disc-plane, with $\theta_{\rm i} = 0^{\circ} (90^{\circ})$ indicating a face-on (edge-on) disc (default $=85^{\circ}$ in text, except stated otherwise). Finally, the angle $\theta_{\rm a}$ is measured in the disc plane from the disc's ascending node on the sky to the positive $x-$axis   (default $=0$ in text, unless stated otherwise). 

Next, we bin the positions of all $N_{\rm t} = N \times N_{\rm p}$ particles in the sky plane with $[X,Y] = [-200, 200]$ AU, with a given resolution of $800\times 800$ pixels. We then compute the disc's surface density $\Sigma$ as projected on the sky by finding the total mass in each bin and dividing it by the bin's area.

Finally, we note that we convolve images of surface density with a 2-D Gaussian having a filter size of $\sim 3 $ AU. In all surface density maps presented throughout the bulk of this paper, (i) the origin is shown by a green star representing the center of mass of the stellar binary and (ii) the logarithmic colour scale represents the surface density distribution normalised by its maximum value. In most of the presented density maps, normalization was performed individually to bring out the  fainter features. 


\bsp	
\label{lastpage}
\end{document}